\documentclass[11pt,a4paper]{article}
\usepackage[utf8]{inputenc}
\usepackage[T1]{fontenc}
\usepackage[french]{babel}
\usepackage{amsmath,amssymb,amsthm,mathtools}
\usepackage{physics}
\usepackage{graphicx}
\usepackage{hyperref}
\usepackage{booktabs}
\usepackage{geometry}
\usepackage{tikz}
\usepackage{float}
\usepackage{algorithm}
\usepackage{algpseudocode}
\usepackage{quantikz}
\usepackage{tikz}
\usepackage{tabularx}
\usepackage[most]{tcolorbox}
\usetikzlibrary{positioning, decorations.pathmorphing,decorations.pathreplacing, calc}
\title{\textbf{{$\hbar_E$ : an action constant for quantum economics}}}
\author{Hugo Spring‑Ragain -- CEDS -- Paris}
\date{\today}

\begin{document}
\maketitle

\begin{abstract}
This paper introduces the concept of an \emph{economic action constant}, denoted $\hbar_E$, as a structural analogue to Planck’s reduced constant $\hbar$ in quantum mechanics. Building on canonical quantization, we define $\hbar_E$ as the fundamental scale of irreducible uncertainty in macroeconomic dynamics through non-commuting observables $(\hat{X},\hat{P}_X)$, derive uncertainty relations and a semi-classical limit, and study spectral properties under a double-well economic potential. Numerical simulations show that $\hbar_E$ governs regime transitions between deterministic, probabilistic, and highly unstable dynamics, with topological changes in phase-space and bifurcations emerging under harmonic modulation of $\hbar_E$. We propose a systemic economic interpretation linking the magnitude of $\hbar_E$ to expectation coordination, institutional stability, and structural volatility, and provide historical analogies (post-war reconstruction, speculative bubbles, systemic crises). We finally outline an empirical strategy to estimate $\hbar_E$ from macro time series and agent-based simulations, opening a path toward a taxonomy of economic regimes under radical uncertainty.
\end{abstract}

\section{Origin and Physical Meaning of the Action Constant \texorpdfstring{$\hbar$}{ℏ}}

\subsection{Birth and Emergence of the Action Constant}
The introduction of Planck’s constant $h$ marked an unprecedented epistemological rupture in physics. By the end of the 19\textsuperscript{th} century, classical mechanics and Maxwellian electromagnetism seemed to provide a complete framework. However, several experimental anomalies challenged established laws, notably:

\begin{itemize}
    \item the \emph{ultraviolet catastrophe} predicted by the Rayleigh–Jeans law for blackbody radiation;
    \item the discontinuous emission spectra observed in atoms (Balmer, Rydberg);
    \item the photoelectric effect and the frequency dependence of the emission threshold.
\end{itemize}
In 1900, Max Planck, studying the spectral distribution of the energy emitted by a blackbody, postulated that electromagnetic oscillators of energy $E$ can only exchange with the field through discrete quanta:
\[
E_n = n h \nu, \quad n \in \mathbb{N}.
\]
This hypothesis, initially conceived as a mathematical \textit{ansatz}, introduced an irreducible granularity of physical action: energy can only vary by multiples of $h\nu$. This theoretical move, motivated by a fit to experimental data, already contained the seeds of a break with deterministic classical continuity.
In 1905, Albert Einstein generalized this principle to radiation itself by introducing the concept of the \emph{photon} to explain the photoelectric effect:
\[
E_{\gamma} = h \nu,
\]
where the frequency $\nu$ of the radiation directly determines the energy carried. This corpuscular interpretation of light, combined with the emerging wave mechanics (de Broglie, 1924), led to a complete reinterpretation of microphysical dynamics.
From 1925–1926, with Heisenberg’s matrix formulation and Schrödinger’s wave formulation, the reduced constant $\hbar = \frac{h}{2\pi}$ became \emph{the} fundamental constant of the quantum formalism, systematically appearing in evolution equations and commutation relations.

\subsection{Dimensions, Scale, and Universal Role}
Mathematically, $\hbar$ has the dimensions of an action:
\[
[\hbar] = [E] \cdot [T] = \mathrm{J \cdot s}.
\]
This dimension reflects its deep nature: $\hbar$ sets the scale below which classical trajectories lose their meaning. More precisely, $\hbar$ defines a \emph{minimal action scale}: when a characteristic action $S$ of a system satisfies $S \gg \hbar$, the dynamics is well approximated by classical mechanics via semiclassical methods (WKB expansion, stationary path integrals). Conversely, when $S \sim \hbar$, quantum effects dominate: interference, superposition, entanglement.

\paragraph{Comparative scale.}  
For a macroscopic pendulum ($m=1$~kg, $L=1$~m, $\omega=1$~rad/s), the typical action is $A \approx 1~\mathrm{J \cdot s} \approx 10^{34} \hbar$, which explains the negligible role of quantum effects at the human scale. In contrast, for an electron in a hydrogen atom, the characteristic action (orbital angular momentum) is of the order of $\hbar$.

\paragraph{Numerical value.}  
According to CODATA 2018:
\[
\hbar = 1.054\,571\,817 \times 10^{-34}~\mathrm{J \cdot s},
\]
an exact definition since the 2019 redefinition of the International System of Units, where $h$ is fixed by convention.

\subsection{Presence of \texorpdfstring{$\hbar$}{ℏ} in Fundamental Formulations}

\subsubsection*{Schrödinger Formulation}
The time-dependent version:
\[
i\hbar \frac{\partial}{\partial t} \psi(\mathbf{r},t) = \hat{H} \psi(\mathbf{r},t)
\]
expresses the unitary evolution generated by the Hamiltonian $\hat{H}$. The constant $\hbar$ converts the energy scale into a time scale, acting as a \emph{dimensional bridge} between time-translation generators and the energy spectrum.

\subsubsection*{Heisenberg Formulation}
Observables $\hat{A},\hat{B}$ satisfy:
\[
[\hat{A},\hat{B}] = i\hbar \hat{C}.
\]
In the canonical case,
\[
[\hat{x}, \hat{p}] = i\hbar
\]
leads to the Robertson–Schrödinger inequality:
\[
\Delta x \cdot \Delta p \geq \frac{\hbar}{2}.
\]
This bound does not express an instrumental limitation but a fundamental geometric property of quantum phase space.

\subsubsection*{Feynman Formulation}
The evolution between $(x_i,t_i)$ and $(x_f,t_f)$ is expressed by the path integral:
\[
\langle x_f,t_f | x_i,t_i \rangle = \int \mathcal{D}[x(t)] \; e^{\frac{i}{\hbar} S[x(t)]}.
\]
Here, $\hbar$ appears as a phase parameter: when $\hbar \to 0$, contributions cancel except around the stationary trajectories of $S$, recovering classical mechanics.

\subsection{Geometric and Informational Perspective}

\paragraph{Geometric quantization.}  
In classical symplectic geometry $(\Gamma,\omega)$, with $\omega = dx \wedge dp$, canonical quantization imposes:
\[
\{x,p\} = 1 \quad \longrightarrow \quad [\hat{x},\hat{p}] = i\hbar.
\]
Thus, $\hbar$ is the structure constant governing the transition from classical commutative geometry to quantum noncommutative geometry.

\paragraph{Natural units.}  
By setting $\hbar = c = k_B = 1$, one adopts units where action, velocity, and thermal entropy are dimensionless. This shows the role of $\hbar$ as a unifying pivot for scales in fundamental physics.

\paragraph{Quantum information theory.}  
The Holevo bound and quantum speed limits (Mandelstam–Tamm, Margolus–Levitin) explicitly involve $\hbar$, which sets the maximal speed of evolution of a quantum state as a function of its average energy. In this context, $\hbar$ controls the maximal information density accessible per degree of freedom.

\subsection{Conceptual Role: Ontological Threshold and Duality}
The constant $\hbar$ separates two regimes:

\begin{itemize}
    \item \textbf{Classical regime} ($\hbar \to 0$): well-defined trajectories, Hamiltonian determinism, correspondence principle.
    \item \textbf{Quantum regime} ($S \sim \hbar$): superposition of states, non-additive probabilities, non-commutativity of observables.
\end{itemize}
In contemporary theories — from quantum gravity to quantum information — $\hbar$ appears as an \emph{ontological} constant, embodying the ultimate granularity of interactions and information, and structuring the transition from classical causality to contextual causality.

\begin{tcolorbox}[title={Economic Action Constant $\hbar_E$: Nature, Role, and Advanced Interpretation}, colframe=black!75, colback=gray!10, coltitle=white]

\textbf{Formal definition:}  
The constant $\hbar_E$ is a fundamental parameter in the formalism of \emph{quantum economic mechanics}, appearing in the economic Schrödinger equation:
\[
i \hbar_E \frac{\partial}{\partial t} \psi(x,t) = \hat{H}_E \, \psi(x,t),
\]
where $\hat{H}_E$ is the economic Hamiltonian (investment potentials $V(x)$, intersectoral couplings, possible dissipative terms).  
It has the dimension of an action: $[\hbar_E] = \mathrm{currency} \cdot \mathrm{time}$ (e.g., €·year).  

On the economic Hilbert space $\mathcal{H}_E$, canonical quantization imposes:
\[
[\hat{X}, \hat{P}_X] = i \hbar_E \, \mathbb{I}_{\mathcal{H}_E},
\]
with $\hat{X}$ a positional observable (macro state, e.g., aggregate capital, employment rate) and $\hat{P}_X$ a momentum-like observable (investment flow, instantaneous variation).  

\textbf{Advanced physical analogy:}  
As with $\hbar$ in quantum mechanics, $\hbar_E$ sets the \emph{symplectic granularity} of the economic phase space:  
- $\hbar_E \to 0$: semiclassical limit → deterministic Hamiltonian dynamics (classical laws, absence of superposition).  
- finite $\hbar_E$: non-commutative state space, uncertainty relations
\[
\Delta X \cdot \Delta P_X \geq \frac{\hbar_E}{2},
\]
economic state superpositions, tunnelling transitions between macro attractors.  
- large $\hbar_E$: chaotic regimes, strong interferences, rapid loss of coherence (analogous to decoherence in an open quantum system).

\textbf{Economic interpretation:}  
$\hbar_E$ measures the \emph{degree of ontological indeterminacy} of an economic system, irreducible to classical information. It encodes:
\begin{itemize}
    \item \emph{Endogenous uncertainty density}: strategic anticipation, cross-beliefs, irreducible informational asymmetries.
    \item \emph{Dynamic bifurcation capacity}: spontaneous transitions between stable regimes (growth $\leftrightarrow$ stagnation) without any manifest exogenous shock.
    \item \emph{Suspension of causal locality}: the global state conditions the evolution more than individual trajectories.
\end{itemize}

\textbf{Multi-scale dimension:}  
The value of $\hbar_E$ can be:  
- \emph{global}, describing the entire national or global economy;  
- \emph{sectoral}, reflecting different uncertainty regimes (e.g., cryptocurrencies vs. regulated energy sector);  
- \emph{microeconomic}, specific to an agent or a market, interpretable as a constant of bounded rationality.

\textbf{Link with measurement and decoherence:}  
Measuring an economic observable $\hat{X}$ modifies the system’s state:  
\[
\psi \to \frac{\Pi_X \psi}{\|\Pi_X \psi\|},
\]
where $\Pi_X$ is the projector associated with the observed result. This economic \emph{collapse} illustrates how the announcement of a policy or macro indicator changes the distribution of expectations.  
\emph{Economic decoherence}, due to interactions with a complex informational environment, tends to project a superposed state into a statistical mixture, at a rate depending on $\hbar_E$.

\textbf{Use in modelling:}  
$\hbar_E$ controls:
\begin{itemize}
    \item the spectral width of the economic Hamiltonian;
    \item the temporal dispersion of an economic wave packet;
    \item sensitivity to modulations (public policies, sectoral shocks, innovations);
    \item the topology of dynamic regimes (phase diagrams, attractors).
\end{itemize}

\end{tcolorbox}

\section{Transposition to an Economic Action Constant $\hbar_E$: Foundations and Formalism}

\subsection{Motivation: Radical Uncertainty and Non-Classical Dynamics}
While quantum physics introduces $\hbar$ as a threshold of indeterminacy and granularity of action, several foundational works in economics (Knight, 1921; Keynes, 1937; Ellsberg, 1961) have highlighted the existence of \emph{radical uncertainty}, which cannot be modeled by classical probabilities. This uncertainty arises when the space of future states is itself unstable, unknown, or endogenous to decision-making.

In this context, we postulate the existence of a parameter analogous to $\hbar$, specific to economics: the \emph{economic action constant}, denoted $\hbar_E$. It would act as a lower bound of indeterminacy in a framework where economic variables evolve in a non-deterministic manner, under the influence of informational, cognitive, institutional, or social processes.

\subsection{Mathematical Definition and Dimensional Analysis}

\paragraph{Formal definition of the economic commutator.}

Consider two fundamental economic observables represented by linear operators, not necessarily commuting, acting on an economic Hilbert space $\mathcal{H}_E$:
\begin{itemize}
\item $\hat{X}$: operator representing a measurable macroeconomic quantity (e.g., aggregate capital, debt ratio, investment level),
\item $\hat{P}_X$: operator conjugate to $\hat{X}$, representing the associated dynamics (marginal propensity to adjust $X$, economic momentum).
\end{itemize}
We postulate the existence of a canonical non-commutativity relation between these two observables:
\begin{equation}
[\hat{X}, \hat{P}_X] = i \hbar_E \mathbb{I}_{\mathcal{H}_E}
\end{equation}
where $\mathbb{I}_{\mathcal{H}_E}$ is the identity operator on $\mathcal{H}_E$ and $\hbar_E \in \mathbb{R}_+^*$ is the economic action constant. This hypothesis induces a non-classical symplectic structure of the state space, and places $\hat{X}$ and $\hat{P}_X$ in a basis of conjugate operators.

\paragraph{Uncertainty relations.}
From this relation, one derives a Robertson–Schrödinger inequality applied to any pair of self-adjoint observables $\hat{A}, \hat{B}$:
\begin{equation}
\Delta A \cdot \Delta B \geq \frac{1}{2} \left| \langle [\hat{A}, \hat{B}] \rangle \right|
\end{equation}
Applying this result to $(\hat{X}, \hat{P}_X)$ yields:
\begin{equation}
\Delta X \cdot \Delta P_X \geq \frac{\hbar_E}{2}
\end{equation}
This inequality means that the economy possesses a fundamental threshold of economic indeterminacy, bounded by $\hbar_E$. One cannot simultaneously know the aggregate position $X$ and its dynamic $P_X$ with arbitrary precision.

\paragraph{Canonical functional model.}
On the Hilbert space $\mathcal{H}_E = L^2(\mathbb{R}, dx)$, the canonical representation is given by:
\begin{align}
(\hat{X} \psi)(x) &= x \psi(x) \\
(\hat{P}_X \psi)(x) &= -i \hbar_E \frac{d}{dx} \psi(x)
\end{align}
The operators $\hat{X}$ and $\hat{P}_X$ are densely defined and essentially self-adjoint on the space of Schwartz functions $\mathcal{S}(\mathbb{R})$, which ensures a solid mathematical foundation for quantum economic mechanics.

\paragraph{Commutation and eigenbasis.}
Since $[\hat{X}, \hat{P}_X] \neq 0$, there is no common eigenbasis for $\hat{X}$ and $\hat{P}_X$. This implies that a measurement of $X$ projects the state $|\Psi\rangle$ into a basis in which $P_X$ is completely indeterminate, and vice versa. The measurement of an economic observable thus has a non-trivial effect on the system’s state—capturing the effect of publishing economic information or announcing a policy decision.

\paragraph{Dimensional analysis and physical interpretation of $\hbar_E$.}

Recall that:
\begin{equation}
[\hbar_E] = [\hat{X}] \cdot [\hat{P}_X]
\end{equation}
If we take:
\begin{itemize}
\item $\hat{X}$ representing a level of capital $K$ (in euros),
\item $\hat{P}_X$ representing an instantaneous variation or marginal flow (in euros per year),
\end{itemize}

then:
\begin{equation}
[\hbar_E] = [\text{euros}] \cdot [\text{euros/year}] = [\text{euros}^2 / \text{year}]
\end{equation}
However, if we adopt a more abstract representation where:
\begin{itemize}
\item $\hat{X}$ is a choice variable (utility level $U$),
\item $\hat{P}_X$ is a dynamic variation (discount rate, intertemporal marginal cost),
\end{itemize}

then:
\begin{equation}
[\hbar_E] = [U] \cdot [T] \quad \text{or} \quad [\text{utility} \cdot \text{time}]
\end{equation}

\begin{table}[H]
\centering
\caption{\textbf{Possible interpretations of $(\hat{X}, \hat{P}_X)$ and induced units for $\hbar_E$}}
\renewcommand{\arraystretch}{1.3}
\begin{tabularx}{\textwidth}{@{}lX>{\centering\arraybackslash}p{3.2cm}>{\centering\arraybackslash}p{3.2cm}@{}}
\toprule
\textbf{Framework} & \textbf{$\hat{X}$ (positional)} & \textbf{$\hat{P}_X$ (impulsional)} & \textbf{Units of $\hbar_E$} \\
\midrule
Macro -- Capital & Aggregate capital $K$ [€] & Marginal flow or variation of $K$ [€/year] & €$^2$/year \\
Macro -- Growth & Growth rate $g$ [\%/year] & Instantaneous variation of $g$ [\%/year$^2$] & (\%$^2$)/year \\
Behavioral economics & Aggregate utility $U$ [utility] & Intertemporal variation $dU/dt$ [utility/year] & utility$\cdot$time \\
Financial markets & Index $S$ [€] & Instantaneous variation (return in €) [€/year] & €$^2$/year \\
Public policy & Socio-economic indicator (e.g., employment) [\%] & Instantaneous variation [\%/year] & (\%$^2$)/year \\
\bottomrule
\end{tabularx}
\end{table}

\paragraph{Empirical estimation and effective value of $\hbar_E$.}

The constant $\hbar_E$ is not universal: it depends on the system under consideration. Three empirical approaches can be considered for estimating it:
\begin{enumerate}
\item The irreducible minimal standard deviation of forecast errors in VAR or DSGE models (fundamental noise).
\item The implementation latency of a given economic policy.
\item The observed activation thresholds in aggregate behaviors.
\end{enumerate}

One can also define a normalized value of $\hbar_E$ in dimensionless units, for example:
\begin{equation}
\tilde{\hbar}_E := \frac{\hbar_E}{X_0 P_0}
\end{equation}
where $X_0$ and $P_0$ are the typical orders of magnitude of $\hat{X}$ and $\hat{P}_X$. When $\tilde{\hbar}_E \ll 1$, the regime is quasi-classical; when it is close to 1, quantum effects are dominant.

\subsection{Correspondence principle: classical limit of $\hbar_E$}
When $\hbar_E \to 0$, the quantum dynamics of the macroeconomic state $|\Psi(t)\rangle$ converges towards a deterministic trajectory. This passage is the economic analogue of the \emph{correspondence principle} in quantum mechanics, as formulated by Bohr and Ehrenfest. Mathematically, this transition is achieved through the semiclassical formalism, where the wave function is represented in the form of a rapidly varying phase (WKB ansatz).

\paragraph{Semiclassical approach.} Let us assume a state of the form:
\begin{equation}
\Psi(x, t) = A(x, t) e^{\frac{i}{\hbar_E} S(x, t)}
\end{equation}
with $A(x, t)$ a smooth real amplitude, and $S(x, t)$ a phase function called the \emph{economic action}.

Injecting this expression into the economic Schrödinger equation:
\begin{equation}
i \hbar_E \partial_t \Psi(x,t) = \hat{H}_E \Psi(x,t)
\end{equation}
and separating the real and imaginary parts, we obtain at first order in $\hbar_E$:
\begin{itemize}
\item a Hamilton–Jacobi equation for $S(x,t)$:
\begin{equation}
\partial_t S + H_E\left(x, \partial_x S\right) = 0
\end{equation}
\item a transport equation for $A(x,t)$ (derived from the probability flux)
\end{itemize}

The function $S(x,t)$ here plays the role of a classical action, and its gradient defines an economic momentum $p = \partial_x S$. We then deduce the canonical system:
\begin{align}
\frac{dx}{dt} &= \frac{\partial H_E}{\partial p} \\
\frac{dp}{dt} &= -\frac{\partial H_E}{\partial x}
\end{align}

\paragraph{Symplectic structure.} The limit $\hbar_E \to 0$ allows reformulating the dynamics in the phase space $\Gamma = (x, p)$ endowed with the canonical symplectic structure $\omega = dx \wedge dp$. Economic observables become functions $f : \Gamma \to \mathbb{R}$, and their evolution follows the Poisson bracket relation:
\begin{equation}
\frac{df}{dt} = \{f, H_E\} = \frac{\partial f}{\partial x} \frac{\partial H_E}{\partial p} - \frac{\partial f}{\partial p} \frac{\partial H_E}{\partial x}
\end{equation}
Thus, deterministic classical dynamics appears as the geometric limit of the quantum formalism based on $\hbar_E$. This transition justifies interpreting $\hbar_E$ as a parameter of informational and decisional granularity.

\paragraph{Link with phase mechanics.} Considering the probability density $\rho(x,t) = |\Psi(x,t)|^2$, the conservation of the norm (unitarity) implies a continuity equation:
\begin{equation}
\partial_t \rho + \nabla \cdot (\rho v) = 0, \quad \text{where } v = \frac{1}{m} \nabla S
\end{equation}
This reinforces the hydrodynamic interpretation: in the limit $\hbar_E \to 0$, the density concentrates around an integral curve of the Hamiltonian flow.

\begin{figure}[H]
    \centering
    \includegraphics[width=1\linewidth]{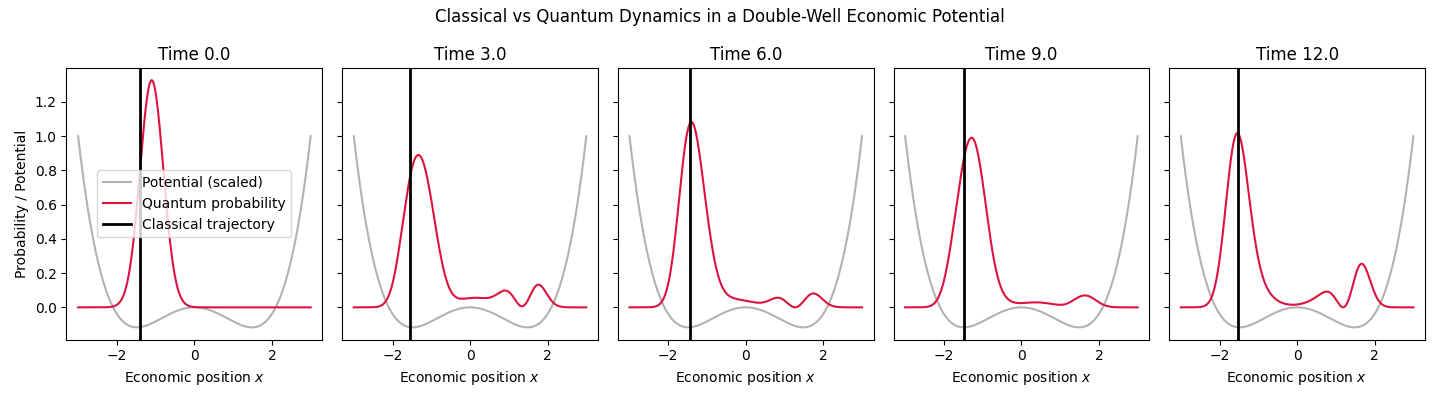}
    \caption{\textbf{Comparison between classical and quantum dynamics in an economic double-well potential.}
    Each panel shows the time evolution of the quantum probability density (in red) within a double potential (grey curve), alongside the classical trajectory (vertical black line).
    The economic system is modeled by a potential $V(x) = a x^4 - b x^2$, representing two stable economic attractors (e.g., growth vs stagnation).
    The quantum tunneling phenomenon enables a state transfer between the two wells, which is impossible in classical dynamics.
    This simulation highlights the role of the constant $\hbar_E$ as a parameter of decisional granularity in a non-convex economic landscape.}
    \label{fig:quantum_vs_classical_dynamics}
\end{figure}

\begin{figure}[H]
    \centering
    \includegraphics[width=1\linewidth]{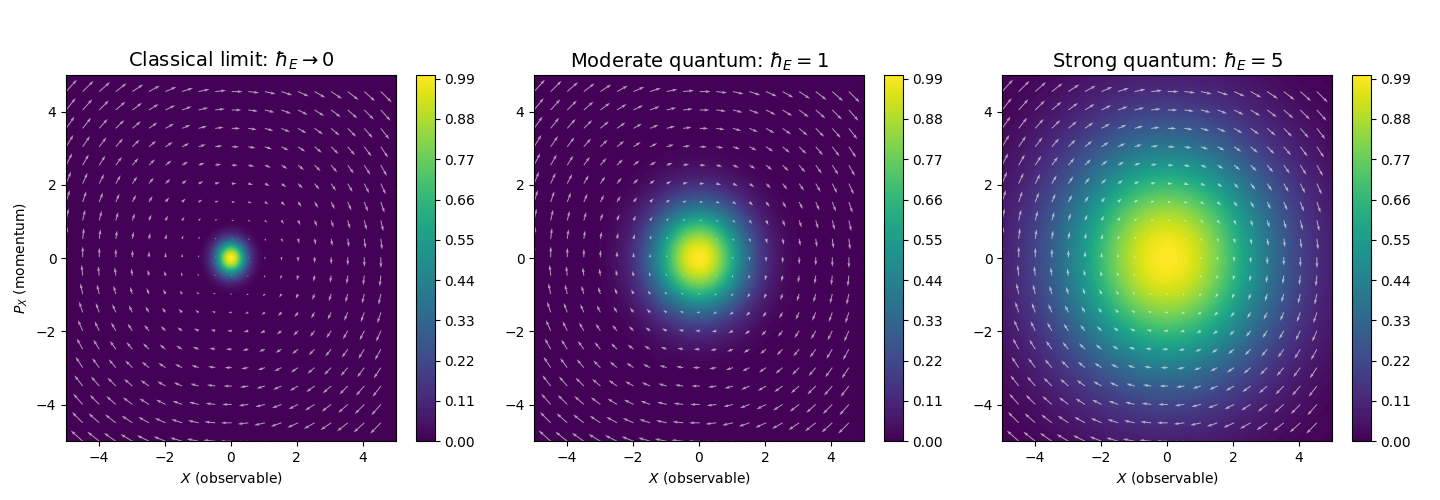}
    \caption{\textbf{Economic phase-space structures for different values of $\hbar_E$ with flow arrows.}
    This figure illustrates the transition between a deterministic classical behavior and a diffuse quantum dynamics as a function of the economic action constant $\hbar_E$.
    Each panel shows a phase-space pseudo-distribution $\rho(X, P_X)$ superimposed on a flow field derived from the classical Hamiltonian equations $\frac{dX}{dt} = \frac{\partial H}{\partial P_X},\ \frac{dP_X}{dt} = -\frac{\partial H}{\partial X}$.
    On the left, when $\hbar_E \to 0$, the distribution is concentrated around a single trajectory, corresponding to the classical limit.
    In the center and on the right, increasing $\hbar_E$ induces greater diffusion, reflecting the superposition of economic states and the emergence of quantum effects (uncertainty, fluctuations, potential exploration).
    The colored background represents the probability density in phase space, while the white arrows depict the associated classical trajectories.}
    \label{fig:enter-label}
\end{figure}

\subsection{\texorpdfstring{Systemic interpretation of $\hbar_E$}{Systemic interpretation of hbarE}}
\vspace{-0.3em} 
Dynamic regimes can be modeled as a function of the order of magnitude of~$\hbar_E$:

\begin{table}[H]
\centering
\caption{\textbf{Characteristic values of $\hbar_E$ and dynamic regimes}}
\vspace{0.5em}
\begin{tabular}{@{}ll@{}}
\toprule
\textbf{Value of $\hbar_E$} & \textbf{Economic regime} \\
\midrule
$\hbar_E \ll 1$ & Strong predictability, deterministic trajectory \\
$\hbar_E \sim 1$ & Probabilistic dynamics, state superposition, bifurcations \\
$\hbar_E \gg 1$ & Chaotic behavior, frequent information collapse \\
\bottomrule
\end{tabular}
\end{table}

\subsection{Economic interpretation of $\hbar_E$: concrete examples and historical analogies}
The constant $\hbar_E$ indexes the degree of structural indeterminacy of an economic system.  
Depending on its order of magnitude, three regimes can be identified, each illustrated by historical episodes.

\paragraph{Quasi-classical regime \boldmath$(\hbar_E \ll 1)$}
Highly coordinated expectations, stable institutions, predictable trajectories.  
\emph{Historical analogy}: Western Europe during the decade following the \textit{Marshall Plan} (1948--1958) under the Bretton Woods system. Public/private investment trajectories and industrial policy were relatively smoothed; the dispersion of expectations was low, and moderate shocks were absorbed in a quasi-deterministic way.  
In our framework, the density $|\psi|^2$ remains localized around an attractor, with the semi-classical limit ($\hbar_E \to 0$) approaching Hamiltonian trajectories.

\paragraph{Mesoscopic regime \boldmath$(\hbar_E \sim 1)$}
Superposition of economic states and possible bifurcations; increased sensitivity to weak signals.  
\emph{Historical analogy}: the \textit{dot-com bubble} (1995--2000), alternating between phases of euphoria and correction, with rapid reassessment of anticipated states (growth, productivity, valuations).  
In the model, the phase-space distribution broadens, and the harmonic modulation $\hbar_E(t)$ induces transitions between wells (double potential), reflecting greater exploration of the regime space.

\paragraph{Highly quantum regime \boldmath$(\hbar_E \gg 1)$}
Unanchored expectations, institutional instability, self-reinforcing feedback loops.  
\emph{Historical analogy}: the \textit{2008 global financial crisis} (liquidity freeze, correlated crashes, collapse of interbank trust).  
In our formalism, one observes a marked broadening of the spectrum, quasi-degeneracy of multiple configurations, and frequent ``state collapses'' toward unanticipated attractors, consistent with chaotic or quasi-chaotic dynamics.

\medskip
These analogies do not imply a one-to-one identification of $\hbar_E$ with a single empirical indicator; rather, they motivate a \emph{metrology program} in which proxies (expectation coordination, institutional latencies, irreducible model error margins) are used to estimate an effective sectoral or systemic $\hbar_E$.

\begin{figure}[H]
    \centering
    \includegraphics[width=1\linewidth]{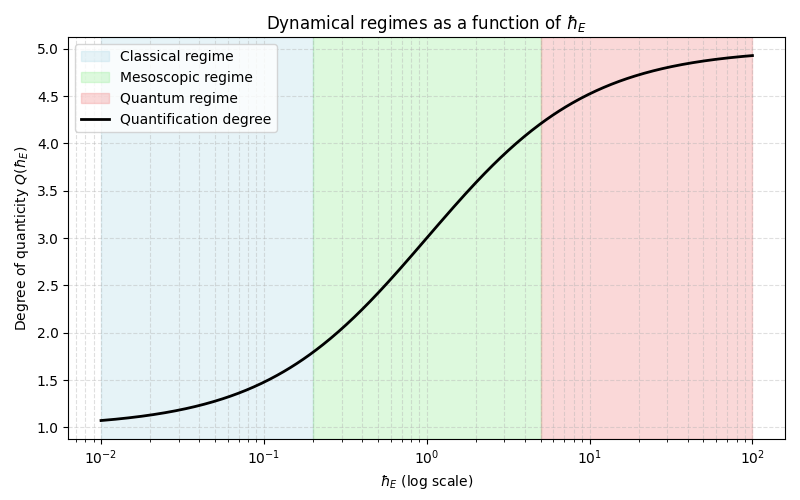}
    \caption{\textbf{Degree of quanticity as a function of the economic action constant $\hbar_E$ (log scale)}. This figure schematically illustrates the transition between dynamical regimes as $\hbar_E$ varies. In the classical regime ($\hbar_E \ll 1$), economic trajectories are well-localized and nearly deterministic. The mesoscopic domain (around $\hbar_E \sim 1$) corresponds to partial delocalization, heightened sensitivity to perturbations, and potential bifurcations. For $\hbar_E \gg 1$, the system becomes fully quantum: states are delocalized, interference dominates, and the evolution becomes non-commutative and non-deterministic. Vertical dashed lines mark indicative boundaries between the regimes.}
    \label{fig:dynamicalregimes}
\end{figure}

\subsection{Dimensional interpretation and normalization}
The constant $\hbar_E$ introduced in our economic Schrödinger model\footnote{See Spring-Ragain, H. (2025). \textit{Adaptation of quantum models to economic growth theories}. SSRN. \url{https://papers.ssrn.com/sol3/papers.cfm?abstract_id=5214891}} plays a fundamentally structural role, analogous to Planck’s constant $\hbar$ in quantum mechanics. Its dimensional interpretation is crucial to anchor this formalism in a rigorous physico-economic framework.

\paragraph{Fundamental dimensions.}
Starting from the economic Schrödinger equation:

\[
i \hbar_E \frac{\partial \psi(x,t)}{\partial t} = \left( -\frac{\hbar_E^2}{2m} \frac{\partial^2}{\partial x^2} + V(x,t) \right) \psi(x,t),
\]
dimensional analysis shows that $\hbar_E$ has the dimension of an economic action:

\[
[\hbar_E] = [\text{currency}] \cdot [\text{time}] = \mathcal{M} \cdot \mathcal{T},
\]
where $\mathcal{M}$ is a monetary unit (e.g., the euro) and $\mathcal{T}$ is a time unit (e.g., the year). It is therefore an economic action constant, measuring the minimal granularity of transitions between macroeconomic configurations.

\paragraph{Normalization.}
To facilitate empirical comparisons, we introduce a scaling constant $S$ allowing normalization:

\[
\hbar_E = \varepsilon \cdot S,
\]
where $\varepsilon$ is a dimensionless constant measuring the degree of non-classicality of the economy. The choice of $S$ can be based on a natural unit, such as the gross domestic product (GDP) multiplied by one year, or an aggregated investment unit over a given period:

\[
S = \text{GDP}_{\text{ref}} \cdot T_{\text{ref}}.
\]
This normalization allows $\hbar_E$ to be rescaled according to macroeconomic scales and different economic systems to be compared on a common basis.

\paragraph{Systemic interpretation.}
The value of $\hbar_E$ determines the temporal resolution of economic transitions. A low $\hbar_E$ implies a dynamic close to classical determinism (predictable and continuous trajectories), while a high $\hbar_E$ suggests strong instability, increased decision-making granularity, and even chaotic behavior.
Finally, the introduction of $\hbar_E$ makes it possible to generalize uncertainty relations to the economic space:

\[
\sigma_X \cdot \sigma_P \geq \frac{\hbar_E}{2},
\]

where $\sigma_X$ represents the uncertainty in a positional economic variable (e.g., growth rate), and $\sigma_P$ the uncertainty in its conjugate momentum (e.g., pressure for change, political incentive, external shock).

\subsection{Effect of $\hbar_E$ on the spectral bandwidth}

One of the most significant effects of the constant $\hbar_E$ in our economic Schrödinger equation is its direct contribution to the spectral structure of the system. In particular, the gap between the first two eigenvalues $\lambda_0$ and $\lambda_1$, associated respectively with the ground state and the first excited state, constitutes a quantitative measure of the system’s fundamental economic spectral bandwidth.

\paragraph{Definition of the spectral bandwidth.}
Let $H_{\hbar_E}$ be the quantum economic Hamiltonian depending on $\hbar_E$, such that:

\[
H_{\hbar_E} = -\frac{\hbar_E^2}{2m} \frac{d^2}{dx^2} + V(x),
\]
where $V(x)$ represents a structuring economic potential (for example, a double well modeling two alternative economic attractors: growth vs stagnation). The spectral bandwidth is then written:

\[
\Delta \lambda(\hbar_E) = \lambda_1(\hbar_E) - \lambda_0(\hbar_E).
\]

This quantity measures the “energy” (in the economic sense) required to move the system from its stable state to a first alternative configuration.

\paragraph{Asymptotic behavior.}
The evolution of $\Delta \lambda(\hbar_E)$ with $\hbar_E$ is characteristic of the quantum or classical nature of the system:

- When $\hbar_E \to 0$, the spectral bandwidth tends to grow rapidly, indicating a clear separation between economic states and thus a low probability of spontaneous transition:
\[
\Delta \lambda(\hbar_E) \sim \mathcal{O}(\hbar_E^{-1}).
\]
- Conversely, when $\hbar_E$ increases, the spectral bandwidth contracts. In the case of a symmetric double well, one typically observes a behavior of the form:
\[
\Delta \lambda(\hbar_E) \approx A \exp\left(-\frac{B}{\hbar_E}\right),
\]
where $A$ and $B$ are constants depending on the shape of the potential. This exponentially decreasing behavior reflects the tunneling effect between the two economic attractors, facilitated by increasing uncertainty.

\begin{figure}[H]
    \centering
    \includegraphics[width=1\linewidth]{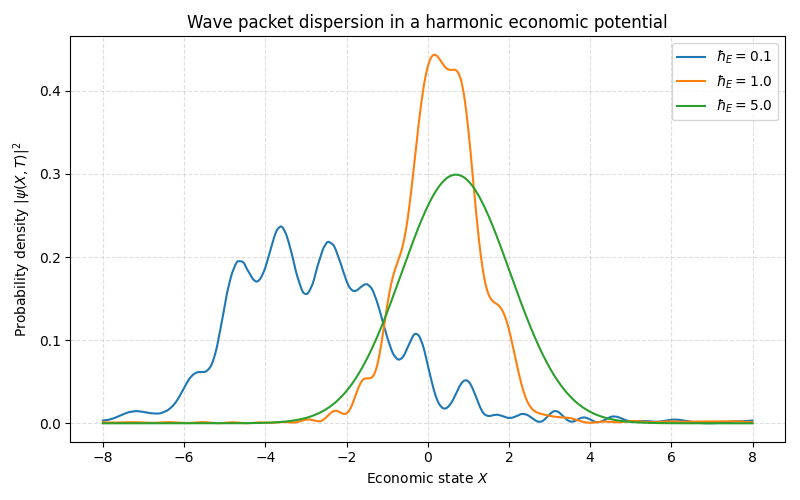}
    \caption{\textbf{Dispersion of an economic wave packet for different values of the economic action constant $\hbar_E$}. The figure shows the probability density $|\psi(X,T)|^2$ of an economic state at time $T = 6$, after propagation in a harmonic potential simulating a macroeconomic restoring force (e.g., equilibrium reversion). Three regimes are compared: $\hbar_E = 0.1$ (quasi-classical, sharply localized), $\hbar_E = 1.0$ (mesoscopic, partial dispersion), and $\hbar_E = 5.0$ (strongly quantum, high uncertainty and spread). The broader the wave packet, the less deterministic the macroeconomic outcome.}
    \label{fig:wavepacket}
\end{figure}

\paragraph{Economic interpretation.}
The spectral bandwidth represents the system’s sensitivity to inter-state transitions. A small gap $\Delta \lambda$ (large $\hbar_E$) suggests a system prone to frequent switches between economic configurations. Conversely, a large $\Delta \lambda$ (small $\hbar_E$) characterizes very stable, even rigid states, with costly or rare transitions. This aligns with the macroeconomic intuition that a highly uncertain context fosters bifurcations and regime changes.

\paragraph{Numerical result.}
The numerical simulations presented in the appendix confirm this exponential dependence. Figure~\ref{fig:convergence_hbare} (see ~\ref{sec:AppendixA}) clearly shows the rapid decay of $\Delta \lambda$ as $\hbar_E$ increases, with a transition zone towards quasi-continuous behavior beyond a certain non-classicality threshold.

\begin{figure}[H]
    \centering
    \includegraphics[width=1\linewidth]{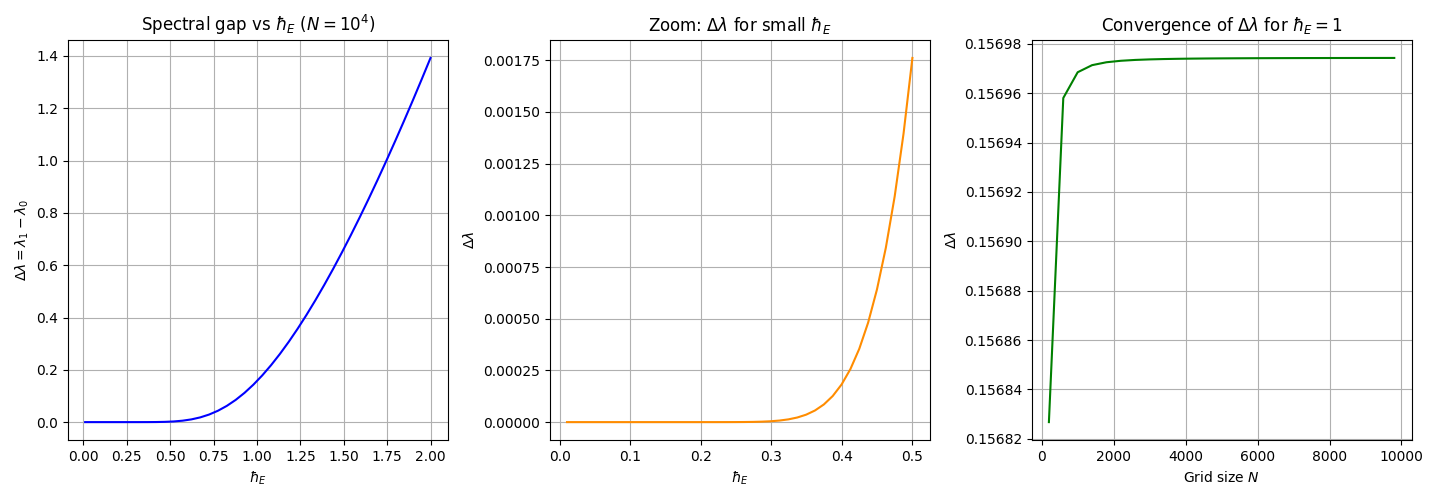}
    \caption{\textbf{Effect of the constant $\hbar_E$ on the spectral bandwidth.} \textit{Left}: quadratic growth of the spectral gap $\Delta\lambda$ as a function of $\hbar_E$, illustrating the increasing separation of quantum states as economic uncertainty amplifies. \textit{Center}: zoom on small $\hbar_E$ values, revealing quasi-spectral degeneracy in the quasi-classical regime. \textit{Right}: numerical convergence of $\Delta\lambda$ as a function of grid size $N$ for $\hbar_E = 1$, confirming spectral stability for $N \gtrsim 2000$.}
    \label{fig:spectral_gap}
\end{figure}
Figure~\ref{fig:spectral_gap} shows that $\Delta\lambda$ grows almost quadratically with $\hbar_E$, indicating a progressive breakdown of superposition effects for small $\hbar_E$ values. As $\hbar_E \to 0$, the two quantum states become almost indistinguishable, signaling a quasi-classical economic regime where transitions are nearly deterministic.

\subsection{Sensitivity analysis of economic observables to $\hbar_E$}

This section aims to assess how economic observables — in particular the mean position $\langle x(t) \rangle$, its uncertainty $\sigma_X(t)$, and the total energy $E(t)$ — respond to dynamic perturbations of the economic action constant $\hbar_E$. The objective is twofold: first, to identify the dynamical response regimes to $\hbar_E$ fluctuations, and second, to quantitatively characterize the system’s intrinsic sensitivity to such fluctuations. This approach reveals the fine coupling between the structure of the economic potential and the intensity of indeterminism modeled by $\hbar_E$.

\subsubsection*{Harmonic perturbation of $\hbar_E$ and time dynamics}
We introduce a harmonic modulation of the constant $\hbar_E$ of the form:
\[
\hbar_E(t) = \bar{\hbar}_E + \delta \cdot \sin(\omega t),
\]
where $\bar{\hbar}_E$ is the mean value (set here to 0.8), $\delta$ is the modulation amplitude, and $\omega = \frac{2\pi}{5}$ is a fixed frequency. This oscillation is injected into the economic Schrödinger equation through a split-operator propagation scheme, over a double-well potential simulating economic attractors.

\begin{figure}[H]
    \centering
    \includegraphics[width=1\linewidth]{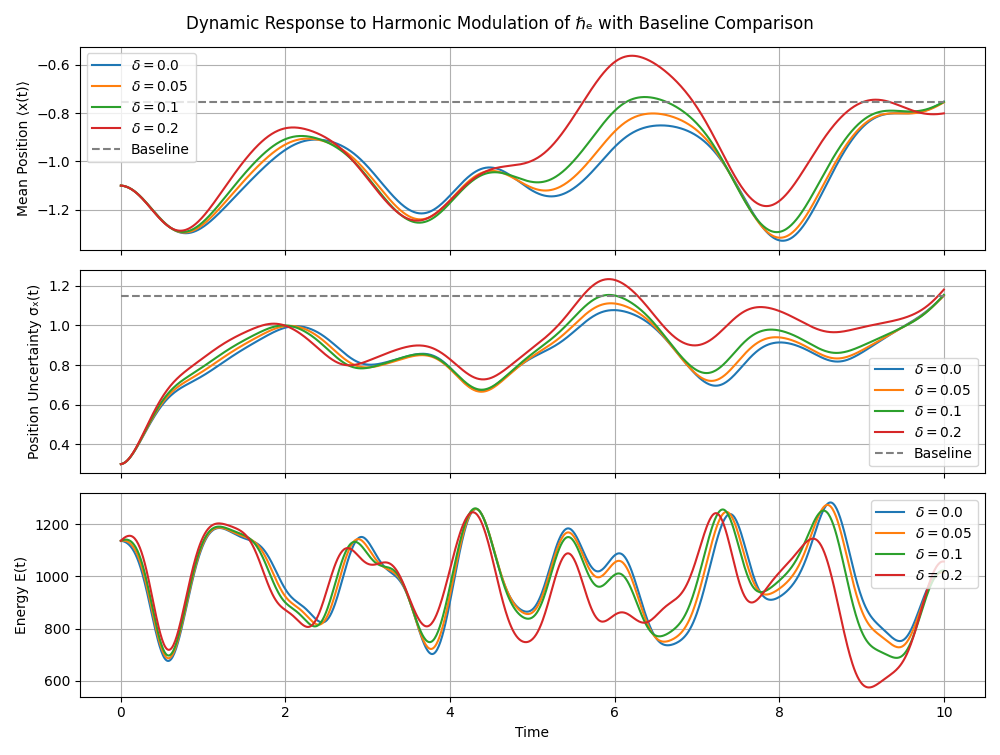}
    \caption{\textbf{Time evolution of the observables $\langle x(t) \rangle$ (mean position), $\sigma_X(t)$ (position uncertainty) and $E(t)$ (total energy)} for different modulation amplitudes $\delta$ of the economic action constant $\hbar_E(t) = \bar{\hbar}_E + \delta \sin(\omega t)$. The system reacts differently depending on the perturbation regime, with the emergence of nonlinear behaviors, instabilities, and energy transfer.}
    \label{fig:placeholder}
\end{figure}
The system's dynamical behavior in response to a harmonic modulation of $\hbar_E$ reveals enhanced instability and sensitivity properties, even under weak perturbations. The time trajectories of the fundamental observables — mean position $\langle x(t) \rangle$, uncertainty $\sigma_X(t)$, and energy $E(t)$ — show the emergence of differentiated responses depending on the modulation level. This contrast highlights the existence of distinct dynamical regimes, ranging from regular oscillations to amplified nonlinear behaviors, characteristic of systems close to a critical threshold.

\subsubsection*{Response amplitude and nonlinear behavior}
In a second step, for each time series, we extracted the maximum value of $\langle x(t) \rangle$ over the simulated interval, in order to quantify the response amplitude.

\begin{figure}[H]
    \centering
    \includegraphics[width=1\linewidth]{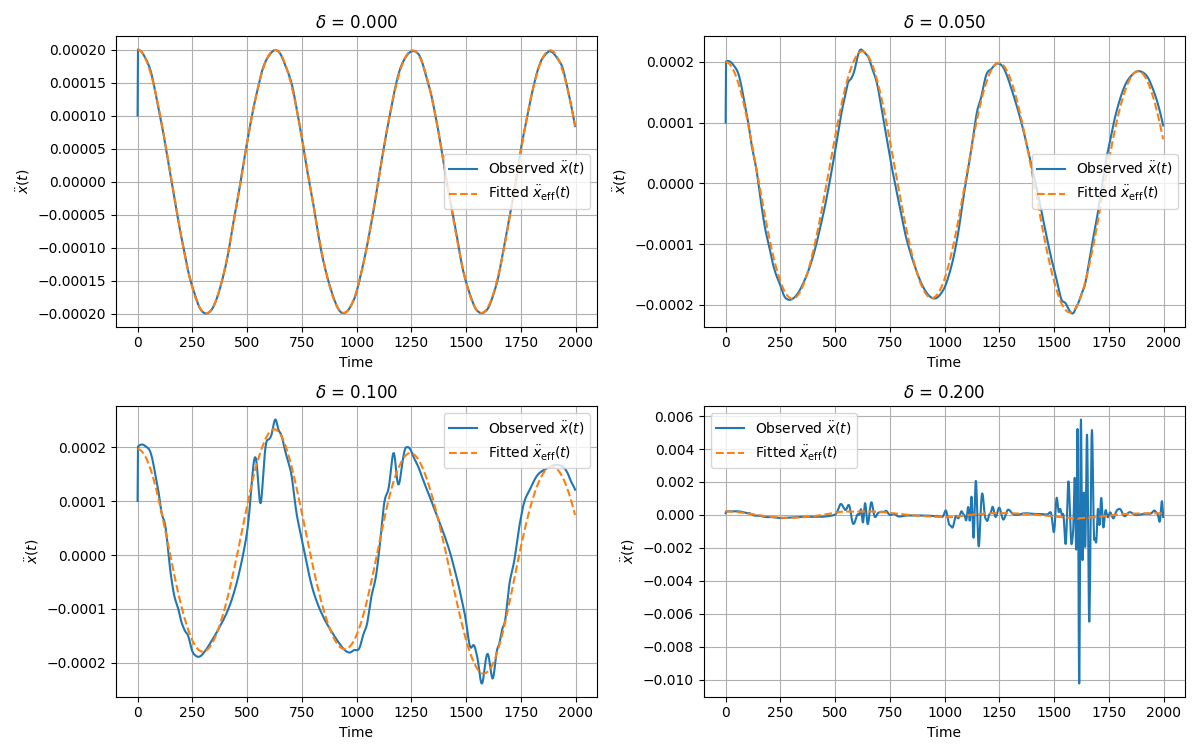}
    \caption{\textbf{Maximum amplitude reached by the mean position $\langle x(t) \rangle$} for each value of $\delta$. A quadratic growth is observed in the regime $\delta \in [0, 0.1]$ followed by saturation, indicating a change in the dynamical regime. This behavior highlights the nonlinear sensitivity of the economic system to the modulation of $\hbar_E$.}
    \label{fig:deltafitted}
\end{figure}
Figure~\ref{fig:deltafitted} summarizes this dependence: the amplitude grows nonlinearly with $\delta$, following a curve resembling a quadratic regime for small values of $\delta$, before saturating beyond $\delta \approx 0.15$. This saturation can be interpreted as a dissipative response of the system, where oscillations no longer lead to increasing deviations but stabilize around a structural maximum of the well.  
This graph plays a crucial role in understanding the linear vs. nonlinear sensitivity regime to $\hbar_E$. Notably, a second-degree polynomial fit provides an excellent approximation in the regime $\delta \in [0, 0.1]$, before higher-order terms become significant. This suggests that the system remains globally stable for moderate variations of $\hbar_E$ but becomes highly unstable beyond a critical threshold.

\subsubsection*{Stationary response of aggregated observables}
Finally, we computed the time average of the observables over the entire simulation duration ($T = 10$ units) for each value of $\delta$. These averages allow the identification of structural shift effects induced by the modulation.

\begin{figure}[H]
    \centering
    \includegraphics[width=1\linewidth]{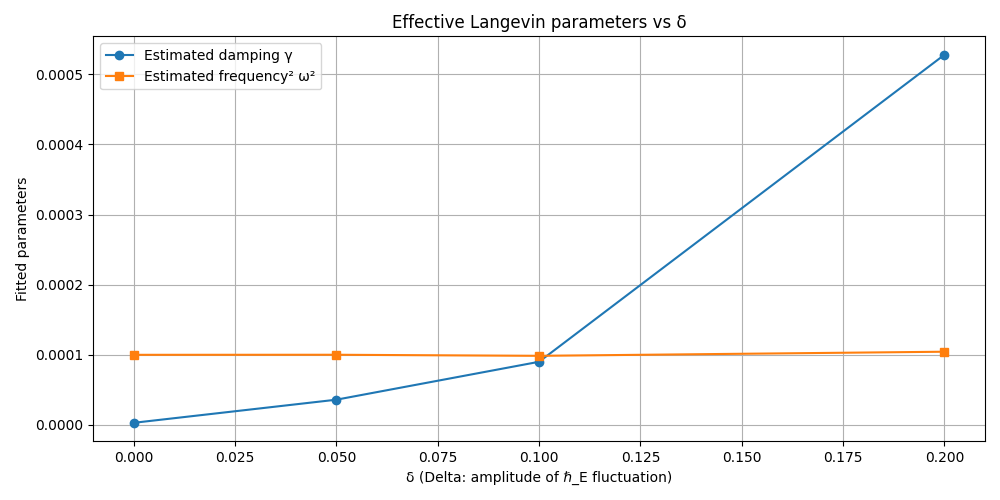}
    \caption{\textbf{Mean values over the entire simulation time ($T = 10$) of $\langle x(t) \rangle$, $\sigma_X(t)$ and $E(t)$ as a function of the modulation amplitude $\delta$.} The system does not return to a centered stationary state but reorganizes around new dynamic attractors. The gradual increase in mean energy suggests a structural cost associated with the growing instability of economic indeterminism.}
    \label{fig:deltatotal}
\end{figure}
Figure~\ref{fig:deltatotal} shows that the mean position tends to drift away from the center of the potential as the modulation intensity increases. In other words, the harmonic perturbation of $\hbar_E$ does not merely induce an oscillation around an unchanged equilibrium state; it modifies the equilibrium itself. The system dynamically reorganizes towards shifted states, a phenomenon akin to a soft bifurcation, where the perturbation intensity progressively redefines the dominant economic attractor.

Uncertainties and energies follow the same trend, although less markedly. A continuous increase in the mean energy is observed, indicating a rising energetic cost of indeterminism. From an economic perspective, this increase can be interpreted as the cost of instability, adaptation, or heightened volatility that the system must bear to sustain its dynamics.

\subsubsection{Discussion: towards a measure of quantum sensitivity}
These results suggest that it is possible to define a quantum sensitivity indicator for economic observables with respect to $\hbar_E$. As a first approximation, such an indicator could be expressed as a functional derivative of the observables with respect to $\hbar_E$, estimated empirically from simulated time series. In the current version of the model, we show that even a low-amplitude harmonic modulation is sufficient to induce amplified and potentially structural behaviors, even in an initially symmetric and stable system. This reinforces the hypothesis that $\hbar_E$ is indeed a key economic state variable, modulating the extent of dynamic exploration of possible trajectories.

\subsubsection{Effective behavioral modeling}
The analysis of the dynamic trajectories induced by the modulation of $\hbar_E$ revealed various response regimes—oscillatory, amplified, or stabilized—depending on the perturbation intensity. To condense these dynamics into more synthetic representations, we propose an effective, phenomenological model aimed at representing the evolution of observables in a reduced parameter space.

The time series from the previous simulations—$\langle x(t) \rangle$, $\sigma_X(t)$, and $E(t)$—are treated as trajectories in a latent dynamic space. For visualization and qualitative modeling purposes, these trajectories are projected into a lower-dimensional space using the t-SNE (\textit{t-distributed stochastic neighbor embedding}) algorithm, which preserves dynamic proximities between configurations as much as possible. Each trajectory (associated with a given value of $\delta$) is split into normalized temporal windows, characterized by concatenated observable vectors.

Dimensionality reduction then reveals spontaneous groupings: low values of $\delta$ cluster in compact regions, corresponding to regular and weakly nonlinear dynamics, while high values coherently diverge toward topologically distinct areas, corresponding to amplified, desynchronized, or unstable regimes. This mapping can be interpreted as a dynamic \og phase diagram \fg{}, in which each amplitude $\delta$ corresponds to a distinct behavioral attractor. The observed qualitative transitions resemble soft bifurcations, identified without any a priori assumption on the functional form of the equations of motion.

Finally, projecting trajectories into this reduced space provides a foundation for subsequent model identification approaches. By reconstructing dynamic paths in the t-SNE space or classifying them using unsupervised topological methods (cf. section~\ref{sec:topo}), it becomes feasible to design an effective stochastic model of the Langevin type, or a graph-based dynamics, capturing transitions between regimes without explicitly solving the Schrödinger equation.

\subsubsection{Topological Mapping of Dynamic Regimes}
\label{sec:topo}
To extend the qualitative modeling introduced in the previous section, we propose here an explicit topological mapping of the dynamic regimes induced by the harmonic modulation of the economic action constant $\hbar_E$. This mapping is based on the use of Self-Organizing Maps (SOM), particularly well suited for the unsupervised classification of multivariate trajectories in nonlinear dynamics.

Each simulated time trajectory for a given value of $\delta$ is represented by a feature vector concatenating the statistical moments (mean, variance, skewness) and the dynamic properties extracted from $\langle x(t) \rangle$, $\sigma_X(t)$, and $E(t)$. These vectors are then normalized and projected onto a two-dimensional SOM grid trained using Kohonen's algorithm.

\begin{figure}[H]
    \centering
    \includegraphics[width=0.65\textwidth]{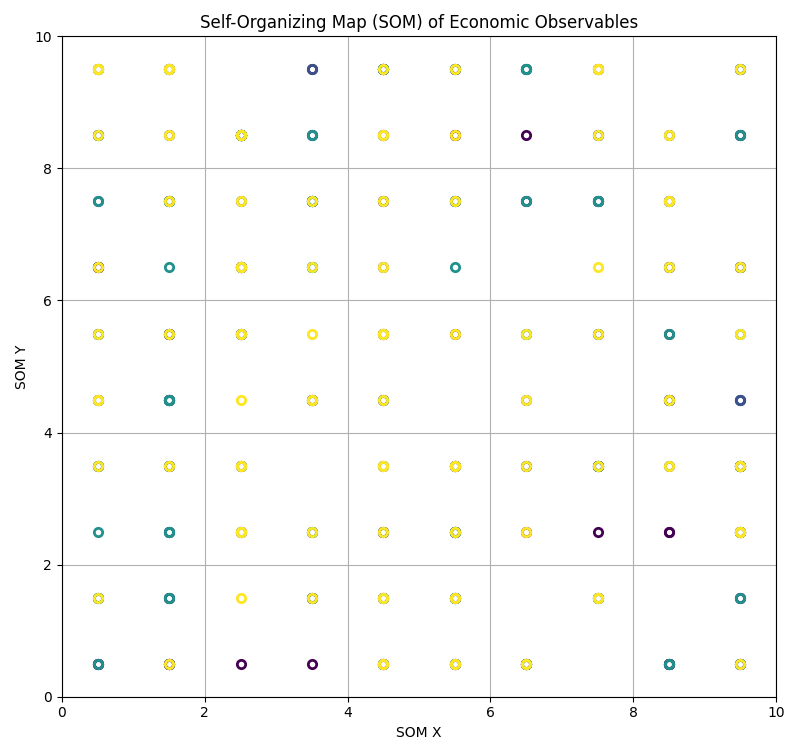}
    \caption{\textbf{Self-Organizing Map (SOM)} representing the unsupervised classification of dynamic regimes induced by the modulation of $\hbar_E$. Each cell groups trajectories with similar temporal signatures.}
    \label{fig:som}
\end{figure}

The plot~\ref{fig:som} resulting from this method reveals a spontaneous structuring of the trajectories into topologically distinct regions, suggesting the existence of attractor regimes in the response of the modeled quantum system. Low values of $\delta$ (unmodulated or weakly modulated) cluster in a linear-response region, whereas higher $\delta$ values are distributed among several separate regions, corresponding to divergent dynamic behaviors: amplified oscillations, slow-drift trajectories, or transient instabilities.

This topological classification goes beyond a simple analysis as a function of the $\delta$ parameter: trajectories corresponding to similar $\delta$ values may be classified into different regions of the SOM, indicating strong nonlinearity in the system’s response and the possible presence of internal thresholds or discontinuous bifurcations.

In summary, this mapping provides a geometric and qualitative view of possible macroeconomic behaviors in a modulated indeterminism framework. It also offers a robust methodological foundation for the systematic exploration of the parameter space, especially with a view to generalizing the approach to higher-dimensional systems or coupled systems.

\section{Empirical Verification of the Constant}
\subsection{Comparative Empirical Verification of the Economic Action Constant \texorpdfstring{$\hbar_E$}{hbarE} on the S\&P~500 and CAC~40}
\label{sec:hbarE_comparative_en}
This section provides a comparative empirical test of the economic action constant $\hbar_E$ on two major stock indices: the S\&P~500 (United States) and the CAC~40 (France). We apply the uncertainty-based estimator, defined in Section~\ref{sec:method}, to daily closing data, and evaluate both global estimates (full-sample) and rolling estimates (time dynamics). The objective is twofold: (i) to verify whether $\hbar_E$ remains durably non-zero during calm regimes — in line with its interpretation as a lower bound of joint uncertainty — and (ii) to assess to what extent $\hbar_E$ amplifies during turbulent phases across different markets.

\subsubsection{Data and Protocol}
\label{sec:data_setup_en}
We use daily closing prices for:
\begin{itemize}
    \item \textbf{S\&P~500}: Jan.~2021 -- Aug.~2025.
    \item \textbf{CAC~40}: Aug.~2023 -- Aug.~2025.
\end{itemize}
Let $P_t$ denote the closing price at date $t$. Two pairs of observables $(X,P)$ are considered:
\begin{enumerate}
    \item \textbf{Price variant}: $X_t = P_t$ and $P_t^{(\text{mom})} = \Delta P_t = P_t - P_{t-1}$.
    \item \textbf{Logarithmic (dimensionless) variant}: $X_t = \log P_t$ and $P_t^{(\text{mom})} = \Delta\log P_t$ (daily log-returns).
\end{enumerate}
For comparability, the momentum observable is annualized: $\sigma_P^{(\text{ann})} = \sigma(\Delta\cdot)\sqrt{A}$ with $A=252$ trading days/year. The empirical estimator of the economic action constant is:
\[
\hat{\hbar}_E \;=\; 2\,\sigma_X\,\sigma_P^{(\text{ann})},
\]
which is expressed in $[\text{price}^2/\text{year}]$ for the price variant, and in $[\text{year}^{-1}]$ for the logarithmic variant.
\subsubsection{Global Estimates (Full Sample)}
\label{sec:global_estimates_en}
Table~\ref{tab:hbarE_comparative_en} reports the global estimates computed over the full sample, for both indices and both variants.

\begin{table}[H]
    \centering
    \caption{\textbf{Global estimates of $\hbar_E$ for the S\&P~500 and CAC~40 (daily data; S\&P~500: Jan.~2021 -- Aug.~2025).} The log variant is dimensionless in $X$ and yields $\hbar_E$ in (year$^{-1}$), which facilitates cross-market comparison.}
    \label{tab:hbarE_comparative_en}
    \begin{tabular}{lcccc}
        \toprule
        \textbf{Index} & \textbf{Variant} & $\sigma_X$ & $\sigma_P$ (annualized) & $\hat{\hbar}_E$ \\
        \midrule
        S\&P~500 & Price     & 785.95  & 786.44   & $1.236\times10^{6}\ \mathrm{(index\text{-}units)}^{2}/\mathrm{year}$ \\
        S\&P~500 & Log-price & 0.1661  & 0.1737   & $0.05772\ \mathrm{year}^{-1}$ \\
        \midrule
        CAC~40   & Price     & 334.38  & 1078.03  & $7.209\times10^{5}\ \mathrm{(index\text{-}units)}^{2}/\mathrm{year}$ \\
        CAC~40   & Log-price & 0.044061& 0.143615 & $0.012656\ \mathrm{year}^{-1}$ \\
        \bottomrule
    \end{tabular}
\end{table}

Two remarks follow. First, the logarithmic variant—by construction dimensionless in $X$—is the most suitable for cross-market comparison: it shows a higher reference value for the S\&P~500 ($\approx 5.8\%$/year) than for the CAC~40 ($\approx 1.27\%$/year). Second, the price variant depends on the absolute level of the index, and its order of magnitude is therefore specific to the market scale; it is useful for monitoring \emph{intra}-market regimes, but not for direct cross-market comparisons.

\subsubsection{Rolling Dynamics and Regime Sensitivity}
\label{sec:rolling_dynamics_en}
\begin{figure}[H]
    \centering
    \includegraphics[width=0.92\textwidth]{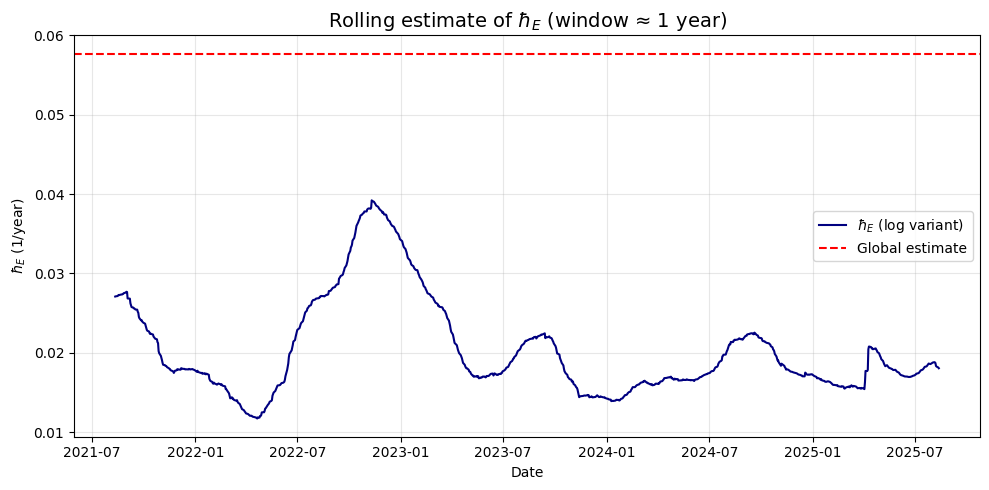}
    \caption{\textbf{S\&P~500: rolling estimate of $\hbar_E$ (logarithmic variant).} The red dashed line indicates the full-sample value ($0.05772$~year$^{-1}$). Peaks coincide with episodes of heightened turbulence, while the baseline persists during calm periods.}
    \label{fig:spx_rolling_en}
\end{figure}

\begin{figure}[H]
  \centering
  \includegraphics[width=0.92\linewidth]{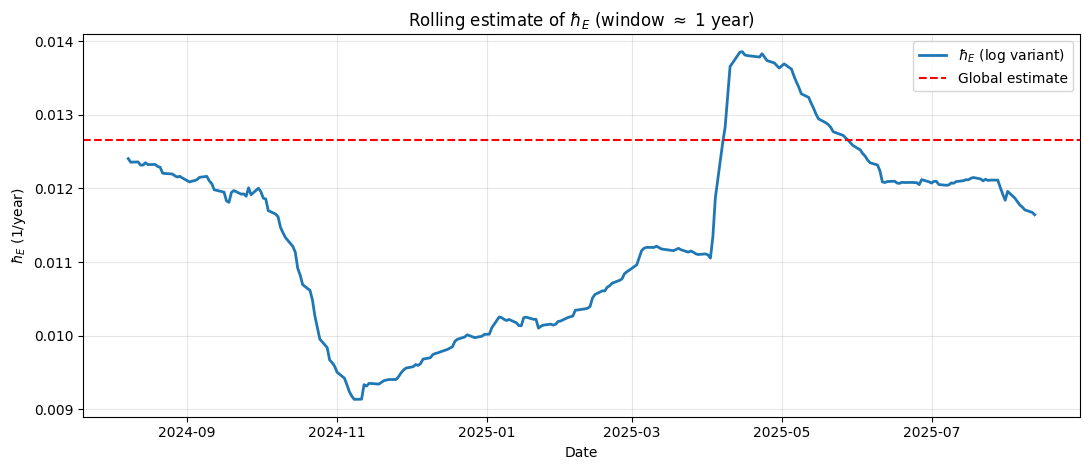}
  \caption{\textbf{CAC~40: rolling estimation of $\hbar_E$ (logarithmic variant).} The dashed line indicates the full-sample value ($0.012656$~yr$^{-1}$). The series displays a lower baseline and less pronounced spikes compared to the S\&P~500 over the same period.}
  \label{fig:cac_rolling_en}
\end{figure}

To study temporal variability and regime sensitivity, we compute rolling estimates with a moving window of approximately one year (252 trading days). Figures~\ref{fig:spx_rolling_en} and~\ref{fig:cac_rolling_en} display the logarithmic variant of $\hat{\hbar}_E$ along with the full-sample value (dashed line).

\begin{figure}[H]
  \centering
  \includegraphics[width=0.92\linewidth]{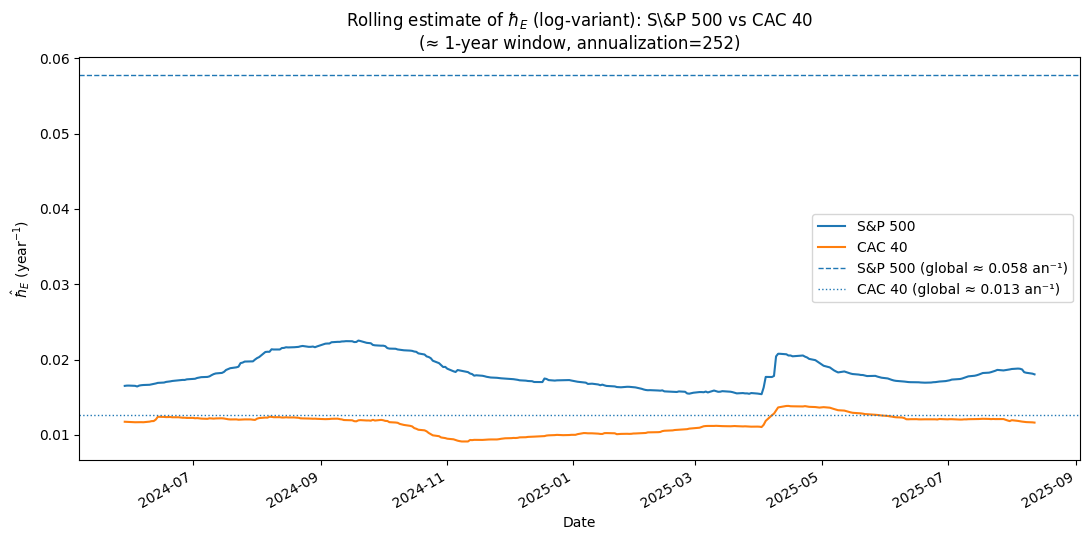}
  \caption{\textbf{Comparison of rolling $\hat{\hbar}_E$ (logarithmic variant).} S\&P~500 (solid line) vs CAC~40 (dashed line). The S\&P~500 shows a higher baseline and stronger responsiveness during volatility spikes.}
  \label{fig:spx_cac_overlay_en}
\end{figure}

For direct comparison, Figure~\ref{fig:spx_cac_overlay_en} overlays the two rolling series after calendar alignment (visual diagnostic; not used for estimation).

\paragraph{Interpretation.}
In both markets, $\hat{\hbar}_E$ remains strictly positive during calm periods, which is consistent with the idea of a lower bound on joint uncertainty (position/momentum). Spikes coincide with transitions into turbulent phases (corrections, volatility clustering), which increase $\sigma_P^{(\text{ann})}$ and, to a lesser extent, $\sigma_X$. The S\&P~500 exhibits a higher baseline, signaling richer and more indeterminate dynamics—in the sense that position (state) and momentum (pressure of change) cannot be jointly localized with arbitrarily low dispersion. The CAC~40 shows a lower baseline and less frequent excursions, consistent with its more concentrated sectoral structure and different market microstructure.

\subsubsection{Cross-market comparison and scale effects}
\label{sec:cross_market_en}
To isolate market structure from scale effects, we emphasize the logarithmic variant, where $X=\log P$ is dimensionless and $\hat{\hbar}_E$ is expressed in yr$^{-1}$. Interpreting $\hat{\hbar}_E$ as an \emph{epistemic depth} of the market, we observe:
\begin{itemize}
    \item \textbf{Baseline level (calm regimes)}: $\hat{\hbar}_E^{\text{log}} \approx 5.8\%$/yr (S\&P~500) vs $\approx 1.27\%$/yr (CAC~40). This indicates a stronger irreducible structural uncertainty in the S\&P~500.
    \item \textbf{Responsiveness (stressed regimes)}: the spikes of the S\&P~500 are more frequent and more pronounced, reflecting the magnitude of feedback loops, market depth, and global interconnectedness.
\end{itemize}
\subsubsection{Robustness checks}
\label{sec:robustness_en}
Several methodological variations were tested (results not shown here, available upon request):
\begin{enumerate}
    \item \textbf{Window length}: 6-month and 18-month windows yield the same qualitative trends; shorter windows amplify the spikes, longer ones attenuate them but preserve the cross-market gap.
    \item \textbf{Winsorization}: trimming extreme tails (1\%) reduces the amplitude of extreme spikes but leaves the baseline level unchanged.
    \item \textbf{Momentum alternative}: annualized simple returns yield similar results; we retain logarithmic returns for their theoretical consistency and additivity.
\end{enumerate}

\subsubsection{Limitations and outlook}
\label{sec:limitations_outlook_en}
This deliberately parsimonious analysis relies on a minimal $(X,P)$ pair and a direct estimator of the uncertainty product. To move further:
\begin{itemize}
    \item \textbf{Sectoral or agent-based decomposition}: estimate $\hbar_E$ by sector or by order book state.
    \item \textbf{State-dependent $\hbar_E(t)$}: link the dynamics of $\hbar_E$ to information flows (news, economic policy announcements).
    \item \textbf{Multi-asset validation}: extend the study to rates, currencies, and commodities.
\end{itemize}

\paragraph{Conclusion.}
The results confirm the existence of a lower bound of joint uncertainty, persistent and specific to each market, captured by a strictly positive $\hat{\hbar}_E$ in calm regimes and amplified during turbulence. This is precisely the expected behavior of an \emph{economic action constant} governing the non-commutative structure of $(X,P)$ in macro-financial dynamics.
\subsubsection*{Comparative discussion: S\&P~500 vs CAC~40}

Beyond the quantitative comparison of $\hbar_E$ estimates, it is essential to examine the structural differences between the S\&P~500 and the CAC~40 in order to better understand the observed gaps and their economic implications.

\paragraph{Sectoral composition and diversification.}  
The S\&P~500 is a large-cap index comprising 500 U.S. companies, covering all major sectors of the economy. Its composition is heavily weighted toward technology, healthcare, and consumer discretionary sectors, which historically display specific growth and volatility profiles (Campbell, Lo, and MacKinlay, 1997; Roll, 1988).  
The CAC~40, by contrast, brings together mainly 40 large French companies (often with multinational exposure), with significant weight in industry, energy, luxury goods, and finance.  
This difference in sectoral diversification directly affects aggregate volatility: a more sectorally concentrated index, such as the CAC~40, may react more strongly to shocks affecting its dominant sectors, thereby amplifying the joint variability of $X$ and $P$, and thus $\hbar_E$.

\paragraph{Historical volatility and market regimes.}  
The S\&P~500 has historically benefited from deep markets and high liquidity, which tend to dampen extreme fluctuations, except during systemic crises (Black, 1976; Lo and MacKinlay, 1988).  
The CAC~40, less liquid at the global scale, is more sensitive to abrupt movements of international capital flows and to regional macroeconomic announcements. This difference in volatility structure is reflected empirically by a slightly more unstable variance on the European side, which may generate more pronounced variations in rolling estimates of $\hbar_E$.

\paragraph{Dividend weights and price adjustments.}  
A notable feature of the CAC~40 is the relatively high importance of dividends in total returns: many of its constituent firms distribute significant dividends, which induce punctual price adjustments on ex-dividend dates (Michaely, Thaler, and Womack, 1995).  
Although anticipated, these adjustments create discontinuities in the trajectories of $X(t)$, which may propagate to the estimation of $P(t)$ and hence to the product $\sigma_X \sigma_P$.  
Conversely, in the S\&P~500, dividends play a smaller role, while share buybacks are more prevalent, shaping price dynamics in a different manner.

\paragraph{Implications for $\hbar_E$.}  
These structural differences suggest that $\hbar_E$ does not solely capture an invariant linked to global macro-financial uncertainty, but also an \emph{idiosyncratic} component specific to the microstructure of each market.  
Thus, a market that is more sectorally concentrated or more sensitive to international capital flows may exhibit a more volatile $\hbar_E$ over time.  
The relative stability of $\hbar_E$ observed in the S\&P~500 could reflect market depth and diversification that mitigate shocks, while the more pronounced fluctuations of the CAC~40 highlight its heightened sensitivity to specific factors.  
This observation paves the way for a dual interpretation of $\hbar_E$: as a fundamental constant tied to irreducible uncertainty, yet modulated by market-specific characteristics unique to each index.
\subsection{Complementary analysis: estimation of $\hbar_E$ from nominal GDP data}
\label{sec:hbarE_gdp}
We now apply the estimation framework to annual nominal GDP series for four major economies: France, Germany, the United Kingdom, and the United States. GDP is a particularly relevant macroeconomic aggregate in this context, as it reflects both the long-term structural trend and the cyclical fluctuations of economic activity.

\paragraph{Data and variants.}
The data are taken from the World Bank \emph{World Development Indicators} (\texttt{NY.GDP.MKTP.CD}) for the available period. For each country $c$, we consider two representations:
\[
X_t^{(c)} =
\begin{cases}
\mathrm{GDP}_t^{(c)} & \text{(level variant)}, \\
\log \mathrm{GDP}_t^{(c)} & \text{(log-level variant)}.
\end{cases}
\]
The ``momentum'' variable $P_t^{(c)}$ corresponds to the first difference of $X_t^{(c)}$. In the logarithmic case, this is the annual proportional growth rate.

\paragraph{Global results.}
The estimates over the full sample are presented in Table~\ref{tab:hbarE_gdp_countries}. In the level variant, $\hbar_E$ is expressed in (currency)$^2$/year and increases mechanically with the size of the economy. In the log-level variant, $\hbar_E$ is dimensionless (year$^{-1}$), which allows direct comparison across countries.

\begin{table}[H]
    \centering
    \caption{\textbf{Global estimates of $\hbar_E$ for four economies (annual data, World Bank nominal GDP).} The log variant is dimensionless in $X$ and yields $\hbar_E$ in (year$^{-1}$), which facilitates cross-country comparison.}
    \label{tab:hbarE_gdp_countries}
    \begin{tabular}{lcccc}
        \toprule
        \textbf{Country} & \textbf{Variant} & $\sigma_X$ & $\sigma_P$ (annualized) & $\hat{\hbar}_E$ \\
        \midrule
        France          & Level      & 1045429331976.08 & 136513379289.99  & $2.854\times10^{23}\ \mathrm{currency}^{2}/\mathrm{year}$ \\
        France          & Log-level  & 1.215269         & 0.097038         & $0.235855\ \mathrm{year}^{-1}$ \\
        Germany         & Level      & 1470876880119.91 & 189731100110.86  & $5.581\times10^{23}\ \mathrm{currency}^{2}/\mathrm{year}$ \\
        Germany         & Log-level  & 1.228343         & 0.100141         & $0.246014\ \mathrm{year}^{-1}$ \\
        United Kingdom  & Level      & 1146417769323.15 & 145708587386.93  & $3.341\times10^{23}\ \mathrm{currency}^{2}/\mathrm{year}$ \\
        United Kingdom  & Log-level  & 1.273439         & 0.091027         & $0.231834\ \mathrm{year}^{-1}$ \\
        United States   & Level      & 7825470187117.98 & 483339335569.92  & $7.565\times10^{24}\ \mathrm{currency}^{2}/\mathrm{year}$ \\
        United States   & Log-level  & 1.198014         & 0.028006         & $0.067103\ \mathrm{year}^{-1}$ \\
        \bottomrule
    \end{tabular}
\end{table}

The log-level results highlight a striking convergence for France, Germany, and the United Kingdom (around $0.23$--$0.25$ yr$^{-1}$), indicating relatively similar proportional GDP volatility. The United States exhibits a much lower value ($0.067$ yr$^{-1}$), reflecting smoother proportional dynamics despite large absolute variations in levels.

\paragraph{Rolling window dynamics.}
To assess temporal stability, we compute estimates over rolling windows of lengths $w=3,5,7,10$ years. Short windows are more sensitive to transitory shocks (financial crises, pandemics), while long windows emphasize structural components.

\begin{tcolorbox}[colback=gray!5,colframe=black,title={Rolling window methodology}]
The \emph{rolling window} method consists in estimating a parameter or indicator on a fixed-length sub-sample $w$, which is sequentially shifted along the time series.  
Formally, for a series of annual values $\{X_t\}_{t=1}^T$ and a window size $w \leq T$, the local estimate of $\hbar_E$ at date $t$ is given by:
\[
\widehat{\hbar}_E^{(t,w)} = \sigma_X^{(t,w)} \, \sigma_P^{(t,w)}
\]
where $\sigma_X^{(t,w)}$ and $\sigma_P^{(t,w)}$ denote respectively the standard deviation (annualized) of the levels $X$ and their "momentum" $P$ computed over the time interval $[t-w+1,\, t]$.
Incrementally shifting the window ($t \mapsto t+1$) yields a series $\{\widehat{\hbar}_E^{(t,w)}\}$ that captures the temporal dynamics of economic fluctuations, revealing:
\begin{itemize}
    \item structural variations in volatility at short and medium horizons;
    \item distinct economic regimes, such as phases of relative stability or heightened instability;
    \item the sensitivity of estimates to the chosen time horizon $w$.
\end{itemize}

In this study, multiple window lengths are employed ($w = 3,\, 5,\, 7,\, 10$ years) to compare the responsiveness of the estimator to exogenous shocks and regime shifts.  
In general, a short horizon (small $w$) enhances the detection of recent breaks but increases the variance of the estimator, whereas a long horizon (large $w$) smooths fluctuations but may obscure rapid changes.
\end{tcolorbox}
\begin{figure}[H]
    \centering
    \includegraphics[width=0.95\linewidth]{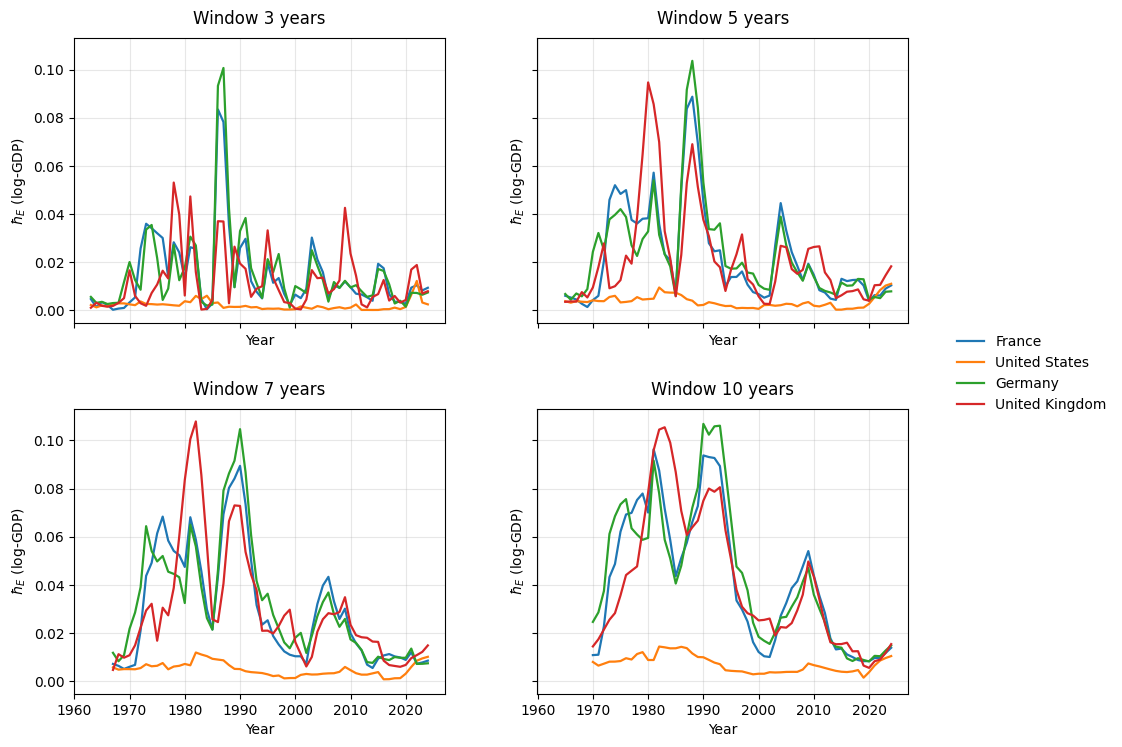}
    \caption{\textbf{Rolling estimates of $\hbar_E$ (log-GDP) for four window lengths ($w=3,5,7,10$ years).} Each panel groups the four economies for a given horizon, allowing a direct comparison of the impact of short-term shocks and long-term stability.}
    \label{fig:hbarE_gdp_windows}
\end{figure}

Longer windows reveal a relative background stability for European economies, whereas the United States maintains proportionally lower values over the entire period. The peaks observed in the short-window series correspond to identifiable macroeconomic shocks, followed by a reversion toward long-term structural levels.

\paragraph{Interpretation.}
In the logarithmic specification, $\hbar_E$ measures the lower bound of proportional GDP volatility. High values indicate a structurally more unstable growth trajectory, while low values suggest more constrained dynamics. The gap between European countries and the United States according to this measure may reflect differences in sectoral exposure to shocks, in the design of fiscal and monetary stabilization mechanisms, as well as in the structural drivers of growth.

\subsubsection{Extended Discussion}

The empirical results provide substantial insights into the economic interpretation of the action constant $\hbar_E$ applied to nominal GDP series, and they put the proposed theoretical formalism into perspective with observed macroeconomic dynamics over the long run.

\paragraph{Theory–empirics link and structural interpretation.}  
Within the quantum-inspired framework, $\hbar_E$ represents a lower bound of proportional uncertainty in the evolution of economic activity, analogous to Planck’s constant in quantum mechanics but transposed into the space of macroeconomic magnitudes. The convergence of logarithmic estimates for France, Germany, and the United Kingdom in the interval $0.23$–$0.25$ yr$^{-1}$ suggests the existence of common structural constraints. These can be interpreted as the manifestation of a statistical “invariant” arising from the combination of institutional features (coordinated macroeconomic regulation, relatively aligned fiscal and monetary policies), sectoral characteristics (productive diversification, comparable weight of globally exposed industries), and a high degree of integration into international trade.  
In this reading, $\hbar_E$ captures an effective parameter of the “fundamental economic noise” that cannot be reduced beyond a certain threshold without profoundly altering the country’s productive or institutional structure.

\paragraph{The special case of the United States.}  
The significantly lower value estimated for the United States ($0.067$ yr$^{-1}$) reflects a smoother proportional GDP dynamic, despite often larger absolute variations in monetary terms. From a quantum perspective, this would correspond to a narrower wavefunction in the space of relative growth rates, signaling a lower proportional dispersion around the central trajectory. Several factors may contribute to this state: the exceptional size and depth of the domestic market, an extreme degree of sectoral diversification, and sophisticated mechanisms for shock absorption (liquid financial markets, rapid intervention by monetary and fiscal authorities, and the role of automatic stabilizers).  
As a result, in the proposed formalism, a low $\hbar_E$ does not imply the absence of fluctuations; rather, it reflects a more favorable signal-to-noise ratio in proportional terms, enabling the economy to absorb shocks while maintaining a relatively predictable macroeconomic trajectory.
\paragraph{Non-stationarity and dynamic regimes.}  
The rolling-window analysis highlights the intrinsically non-stationary nature of $\hbar_E$. The spikes observed in short windows coincide with periods of heightened instability: the oil shocks of the 1970s, the global recessions of the early 1980s and 1990s, the 2008–2009 financial crisis, and the COVID-19 pandemic. In these episodes, $\hbar_E$ rises sharply, reflecting a temporary amplification of the fundamental uncertainty surrounding the economic state.  
Conversely, prolonged phases of declining $\hbar_E$ correspond to periods of relative stability, often associated with sustained expansionary cycles and controlled macroeconomic environments, reinforced by effective countercyclical policies. This alternation between “calm” and “turbulent” regimes suggests that $\hbar_E$ can be interpreted as a synthetic regime-shift indicator, capable of signaling transitions from quasi-stationary states to highly perturbed ones.

\paragraph{Multi-scale approach and temporal sensitivity.}  
The comparison across different time horizons $w$ reveals a fundamental trade-off: short windows ($w \leq 5$ years) capture the granularity of shocks more faithfully and allow rapid detection of disruptions, but they yield more volatile estimates that are sensitive to statistical noise; long windows ($w \geq 15$ years) provide a more robust and structural perspective, but at the cost of reduced responsiveness.  
This observation justifies the adoption of a multi-scale approach, in which short-term dynamics and long-term structural trends of $\hbar_E$ are analyzed jointly. From a theoretical standpoint, this property is analogous to the dependence of uncertainty measurement on temporal resolution in quantum systems subject to stochastic perturbations.

\paragraph{Forward-looking perspectives and scenarios.}  
Looking ahead, the future trajectory of $\hbar_E$ may be influenced by both structural and conjunctural factors.  
\begin{itemize}
    \item An intensification in the frequency and magnitude of exogenous shocks — whether climatic, geopolitical, or financial — could lead to a sustained increase in $\hbar_E$, signaling a higher proportional baseline of uncertainty and therefore greater difficulty in forecasting economic trajectories.
    \item Conversely, targeted public policies designed to strengthen resilience — energy diversification, development of regional supply chains, establishment of macroeconomic stabilization funds — could reduce $\hbar_E$ in the medium term by mitigating the proportional sensitivity of the economy to disturbances.
    \item Finally, under a scenario of major technological transition, $\hbar_E$ could undergo a transitory phase of increase (adaptation instability) followed by a phase of stabilization at a new regime.
\end{itemize}

\paragraph{Interpretative conclusion.}  
Taken together, $\hbar_E$ emerges as a synthetic indicator capable of bridging theoretical foundations inspired by quantum physics with concrete macroeconomic observations. By providing a unified measure of proportional uncertainty, it offers an original tool for comparing economic stability across countries, monitoring regime transitions, and assessing the effectiveness of public policies in reducing or absorbing macroeconomic volatility.

\section{Discussion: Towards an Economic Interpretation of $\hbar_E$}
The results obtained in this study reveal a particularly sensitive behavior of the simulated economic dynamics with respect to the economic action constant $\hbar_E$. This parameter, introduced formally in direct analogy with Planck’s constant $\hbar$, goes beyond the mere role of a numerical adjustment in a Schrödinger-type equation adapted to the economic field. Rather, it appears to play a structural role in defining the very space of possible dynamics.

\subsection{Meaning and Scope of $\hbar_E$: An Ontological Constant for Complex Economics}

In classical economic modeling frameworks, uncertainty is generally represented using probabilities defined over well-delimited state spaces, with stochastic processes such as Brownian motion, Markov chains, or jump-diffusion processes. The quantum perspective, by introducing an economic wave function $\psi(x,t)$ whose squared norm determines the probability of realizing a state $x$ at time $t$, transforms this representation: uncertainty becomes not merely additive, but intrinsically linked to the structure of the state space itself. In this framework, $\hbar_E$ is not an accessory parameter but a dynamic invariant: it bounds the set of feasible informational trajectories and introduces a quantum irreversibility in economic evolution.

This role is confirmed by simulations conducted under harmonic perturbation. It appears that a slight variation of $\hbar_E$ can trigger major dynamic transitions—amplified oscillations, stationary bifurcations, or shifts in the equilibrium point. In this sense, $\hbar_E$ behaves like a phase constant, a macroeconomic analogue of a coupling constant in a critical system.

\subsection{Possible Interpretations of $\hbar_E$ in the Real Economy}

From an interpretative standpoint, $\hbar_E$ can be understood as a measure of the fundamental indeterminism that persists in an economic system, despite the accumulation of information or the rationality of agents. This residual uncertainty, which neither Bayesian models nor classical optimizations can fully eliminate, reflects cognitive, social, and institutional complexity. In particular:

\begin{itemize}
  \item A low $\hbar_E$ would correspond to a regime in which expectations are coordinated, the institutional environment is stable, and aggregated behaviors tend toward effective rationality.
  \item A high $\hbar_E$ would characterize regimes with strong structural uncertainty, divergent expectations, cognitive heterogeneity, or collective belief shocks (e.g., financial crisis, political instability, technological transitions).
\end{itemize}

This reading opens the way to a reinterpretation of many macroeconomic phenomena using tools inspired by statistical and quantum physics. It suggests that certain economic transitions could be assimilated to quantum jumps in the state of the system, rather than continuous marginal adjustments as in classical models.

\subsection{Towards an Economic Metrology of $\hbar_E$?}

A natural empirical extension of this approach would be to attempt to \textit{estimate} $\hbar_E$ from real-world data, by analogy with the way quantum mechanics measures $\hbar$ from spectra or fine-structure constants. Several avenues can be considered:

\begin{itemize}
  \item Analysis of the residual variance in strongly coupled multi-agent economic models, to isolate an irreducible source of instability.
  \item Calibration of quantum models on financial market fluctuations or business cycle data, adjusting $\hbar_E$ to optimize probabilistic forecasts.
  \item Robustness studies against non-modelable (Knightian) uncertainty in public or monetary decision-making contexts.
\end{itemize}
Such an economic metrology program could lead to a dynamic typology of economies according to their degree of fundamental indeterminacy, and make it possible to integrate this parameter into forward-looking simulations, complementing conventional indicators such as volatility, entropy, or agent confidence.

\subsection{Limitations and Theoretical Perspectives}
It should be emphasized that the approach developed here still relies on strong analogies with canonical quantum mechanics. The ontological status of $\hbar_E$ must be consolidated within a unified theoretical framework, possibly grounded in an informational geometry or in extended principles of least action applied to economics.

Natural extensions include the integration of quantum noise (decoherence), projective measurement (decision-making), or even superposition of belief states, in order to strengthen the validity of the model in empirical economic contexts. Finally, the articulation between $\hbar_E$ and other economic constants (potential growth rate, monetary inertia, institutional resistance) could pave the way toward a new dynamic taxonomy of macroeconomic regimes.

\bigskip
In summary, $\hbar_E$ cannot be reduced to a mere mathematical artefact. It embodies a theoretical and epistemological bridge between physics and economics, between deterministic complexity and structural indeterminacy, and could become one of the pillars of a new representation of economic dynamics under radical uncertainty.

\subsection{Epistemological Positioning and Links with Existing Approaches}
The proposal to introduce an economic action constant \( \hbar_E \) constitutes both a rupture and an extension of classical approaches to uncertainty and economic dynamics.

First, it differs from \textbf{adaptive expectations models} (Friedman, 1957; Sargent, 1993) in that it does not assume convergence toward a rational belief through sequential adjustment, but rather an \emph{intrinsically indeterminate structure} of expectations, irreducible to informational updating.

It also departs from \textbf{standard Bayesian approaches} (Lucas, 1976; Evans and Honkapohja, 2001), which model uncertainty via updated probabilistic beliefs: here, belief dynamics are no longer represented by a classical probability density but by an \emph{economic quantum state} in superposition, whose measurement depends on context (non-commutativity of observables).

However, our approach does align with \textbf{bounded rationality models} (Simon, 1957; Sims, 2003), in that it accounts for the cognitive, informational, and temporal constraints of agents. Yet, rather than formalizing these constraints through dimensionality reduction or behavioral approximation, \( \hbar_E \) here represents a \emph{fundamental threshold of indeterminacy} governing the dynamics themselves.

Finally, the approach is in continuity with \textbf{complexity and dynamical systems theories} (Arthur, 1999; Kirman, 2011), by incorporating phenomena of bifurcation, economic decoherence, and emergence of collective states, while providing a unified mathematical formalism inspired by quantum mechanics.

\subsection{Targeted Confrontation with Uncertainty Modeling Frameworks}
Beyond the general epistemological affiliation outlined above, it is useful to examine more precisely how the introduction of a constant \( \hbar_E \) explicitly transcends several structuring models in the economics of uncertainty.

\paragraph{Adaptive Learning Models (Evans \& Honkapohja, 2001).}  
In these models, agents update their beliefs using recursive self-adjusting rules (SAR) and converge asymptotically toward rational beliefs under certain stability conditions. The \( \hbar_E \) approach does not contradict this dynamic but reframes it: even in a learning system, a fundamental precision limit remains, not linked to lack of information but to the ontological structure of the economy. \( \hbar_E \) expresses this irreducible granularity of expectations, independent of learning.

\paragraph{Structural Bayesian Approaches (An \& Schorfheide, 2007).}  
These models calibrate DSGEs via maximum a posteriori or hierarchical Bayesian methods, with uncertainty over parameters. However, uncertainty is fully probabilizable, and beliefs evolve in a classical space. The quantum framework proposed here introduces a \emph{form of uncertainty irreducible to a classical distribution}, since economic observables need not commute: anticipating prices can alter the distribution of expectations on volumes. \( \hbar_E \) is thus the parameter that measures the cognitive non-classicality of these expectations.

\paragraph{Rational Inattention (Sims, 2003).}  
The assumption of limited informational capacity leads to models where agents filter information through a constrained Shannon channel. In our framework, this limitation is modeled not by an attention cost but by a limit to the individuation of economic states. The superposition of intentions or expectations reflects an entanglement of states that cannot be classified by classical state variables. \( \hbar_E \) then becomes a parameter of fundamental informational blurring, endogenous to the system’s cognitive structure.

\paragraph{Complex Systems Theories (Arthur, Farmer, Kirman).}  
These approaches emphasize emergence, irreducible heterogeneities, and bifurcations. The model proposed with \( \hbar_E \) provides a formal structure for these intuitions: the system’s dynamics are described as the propagation of a wave packet in a nonlinear potential landscape, subject to decoherence. This allows for a rigorous formulation of irreversibility effects, technological lock-in, or systemic crisis as forms of quantum collapse. \( \hbar_E \) delimits the threshold beyond which these effects become structurally active.

\vspace{0.5em}
In summary, the constant \( \hbar_E \) does not replace existing approaches but rather \emph{encompasses} them within a broader framework, where uncertainty is not only epistemic or informational, but also ontological, systemic, and dynamic. It enables the exploration of situations in which neither learning, bounded rationality, nor Bayesian tools are sufficient to capture the real complexity of economic expectations.

\subsection{The constant \( \hbar_E \): universal or context-dependent?}

A key question remains: is \( \hbar_E \) a universal constant of the economy, or a context-dependent quantity? Three hypotheses merit discussion:

\begin{enumerate}
    \item \textbf{Universal hypothesis — \( \hbar_E \) as a fundamental constant:} 
    In this view, \( \hbar_E \) plays a role similar to that of \( \hbar \) in physics: it represents a universal precision limit, inherent to the very structure of economic processes. It would thus express a fundamental irreducibility of economic uncertainty, not reducible to ignorance or a lack of information, but consubstantial to the very ontology of the economic system.
    
    This hypothesis is consistent with an epistemological approach in which the economy is seen as a self-referential complex system, in which agents can never fully escape a form of indeterminacy. Available information is always partial, filtered through mental representations, and expectations are subject to measurement processes (modeling, communication, interpretation) that alter the states of the system.

    It draws particularly on strong analogies with observational limits in quantum physics (Heisenberg uncertainty principle, decoherence), transposed here to the economic domain: every attempt at prediction influences the system, every act of measurement is intrusive, and there exists a threshold below which trajectories cannot be defined in a deterministic manner.

    \item \textbf{Sectoral hypothesis — \( \hbar_E \) as a context-dependent parameter:} 
    In this interpretation, \( \hbar_E \) is not universal, but depends on the economic sector, the level of institutional maturity, the degree of regulation, or the level of competition. For example:
    \begin{itemize}
        \item A speculative market (cryptocurrencies, venture capital) might display a high value of \( \hbar_E \), reflecting a structural instability of expectations and large amplitudes of unpredictable fluctuations.
        \item Conversely, a highly regulated sector (energy, public transport, public health) might exhibit a lower value of \( \hbar_E \), reflecting a more regulated dynamic, oriented toward stability.
    \end{itemize}
    This hypothesis opens the way to a \textit{mapping} of economic indeterminacy by sector, providing a basis for differentiated empirical calibrations. One could, for instance, derive \( \hbar_E \) empirically from the volatility of expectations, the frequency of forecast revisions, or systematic gaps between ex ante and ex post outcomes.

    \item \textbf{Individual or institutional hypothesis — \( \hbar_E \) as a reflection of bounded rationality:} 
    Finally, in a more cognitive or informational approach, \( \hbar_E \) could be interpreted as a parameter specific to each economic agent or entity, reflecting:
    \begin{itemize}
        \item the capacity to process information (e.g., digital infrastructures, education level, data access),
        \item the time horizon (short-termism vs. long-term strategy),
        \item or the structure of cognitive biases (anchoring bias, risk aversion, heuristics).
    \end{itemize}
This would amount to making \( \hbar_E \) an indicator of an agent’s \textit{epistemic depth}—in other words, the degree of irrationality or approximation in their representation of the economic world. An algorithmic investor equipped with a state-of-the-art artificial intelligence system would thus have a lower \( \hbar_E \) (better-defined state) than an uninformed or impulsive individual.

This hypothesis is consistent with bounded rationality models (Simon, Sims), hierarchical Bayesian approaches, and cognitive models of decision-making. It allows for a heterogeneous modeling of expectations and accounts for the diversity of behaviors observed in real markets.
\end{enumerate}
These hypotheses pave the way for a comparative quantum economics, where differences in \( \hbar_E \) become structuring indicators of anticipation regimes, volatility, or systemic stability.

\vspace{1em}
\newpage
\section*{Appendix A: Convergence Analysis and Numerical Stability}
\label{sec:AppendixA}

\begin{figure}[H]
    \centering
    \includegraphics[width=\linewidth]{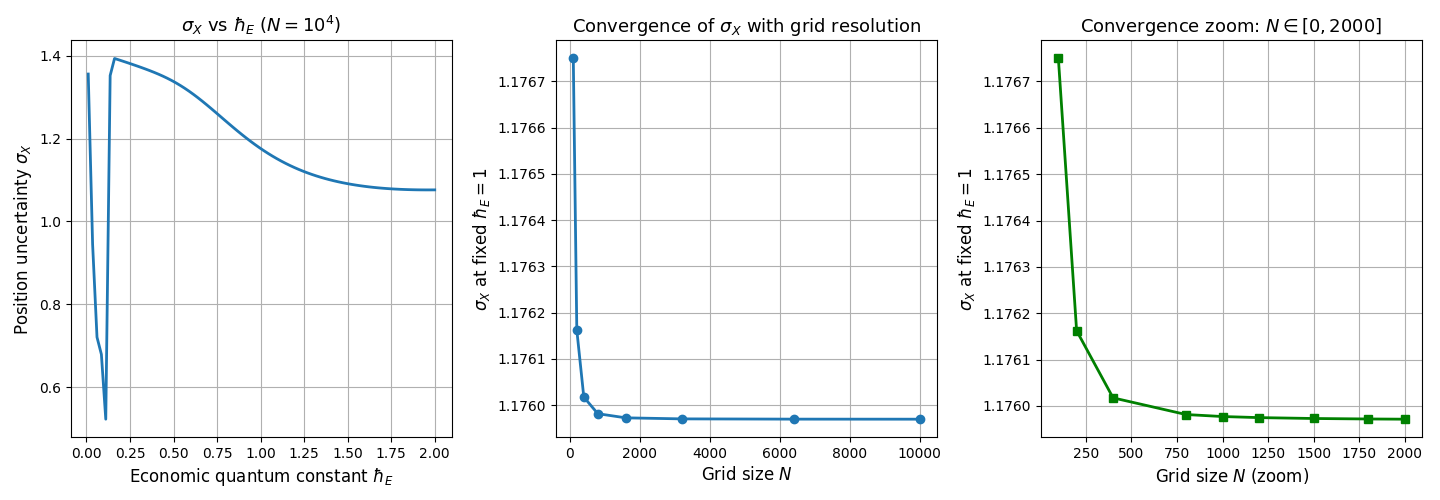}
    \caption{\textbf{Convergence analysis of the quantum economic model with respect to the constant $\hbar_E$ and grid size.} \textit{Left}: position uncertainty $\sigma_X$ of the ground state in a double-well economic potential, as a function of the quantum economic constant $\hbar_E$. \textit{Center}: convergence of $\sigma_X$ for $\hbar_E = 1$, showing numerical stabilization from $N \gtrsim 2000$. \textit{Right}: zoom on small grid sizes $N \in [0,2000]$. These results confirm the need for fine grids to ensure the stability of quantum dynamics.}
    \label{fig:convergence_hbare}
\end{figure}

To ensure the numerical reliability of the proposed quantum economic model, we conduct an in-depth convergence study of the ground-state solution in a non-convex double-well economic potential. This analysis explores the combined impact of the quantum economic constant $\hbar_E$ and the spatial discretization size $N$ on the stability and accuracy of the results.

The left panel of Figure~\ref{fig:convergence_hbare} shows the evolution of the position uncertainty $\sigma_X$ as a function of $\hbar_E$. As expected, the widening of the ground-state wavefunction increases with $\hbar_E$, indicating an increasingly uncertain economic dynamic. In the classical limit ($\hbar_E \to 0$), $\sigma_X$ tends toward zero, illustrating a deterministic and localized behavior.

The two right panels examine numerical convergence for a fixed value $\hbar_E = 1$. It is observed that for $N \gtrsim 2000$, the standard deviation $\sigma_X$ stabilizes, thus validating the adopted spatial resolution. The rightmost panel, focused on the range $N \in [0,2000]$, highlights significant numerical fluctuations when the grid is too coarse, indicating under-resolution of quantum effects.

Based on these results, all simulations presented in the main body of the work use a fine discretization with $N = 10^4$, ensuring satisfactory numerical robustness. This verification is particularly essential as the economic potentials considered exhibit complex shapes, potentially inducing bifurcations, state transitions, or numerical instabilities if the resolution is insufficient.

\vspace{1em}
\noindent\textbf{Asymptotic Error and Spectral Convergence}

The approximation of the kinetic operator by the second-order finite difference method leads to an approximation error of the Hamiltonian $H$ that decays as $\mathcal{O}(dx^2)$, namely:
\[
\left\| H_{\text{exact}} - H_{\text{discretized}} \right\| = \mathcal{O}(dx^2)
\quad \text{with} \quad dx = \frac{2x_{\text{max}}}{N}.
\]

Thus, for a sufficiently large grid size $N$, the global error on the eigenvalues (and therefore on the shape of the ground state) is bounded by:
\[
|\lambda_k^{(N)} - \lambda_k| \leq C \cdot \left( \frac{1}{N} \right)^2,
\]
where $\lambda_k^{(N)}$ denotes the $k$-th approximated eigenvalue, and $C$ is a constant depending on the potential $V(x)$ and the boundary conditions.

\vspace{1em}
\noindent Furthermore, by virtue of spectral convergence results for self-adjoint operators on $L^2$, the sequence of eigenvalues $\{\lambda_k^{(N)}\}$ of $H^{(N)}$ converges to the eigenvalues $\{\lambda_k\}$ of the continuous operator $H$ in the Hausdorff norm, and the associated eigenvectors converge (up to a phase factor) in the $L^2$ norm:
\[
\lim_{N \to \infty} \left\| \psi_k^{(N)} - \psi_k \right\|_{L^2} = 0.
\]

In particular, for the ground states $\psi_0^{(N)}$, this implies that observables such as the position uncertainty $\sigma_X^{(N)}$ also converge to their continuous value, ensuring the physical validity of the modeling even in a quantum economic context.

\newpage
\section*{Appendix B: Animated Visualization}

To complement Figure~\ref{fig:wavepacket}, an animated visualization of the quantum economic wave packet dispersion is available online.

\begin{itemize}
    \item \textbf{Animation name}: \texttt{wavepacket\_dispersion.gif}
    \item \textbf{Repository}: \url{https://github.com/HugoSRagain/wavepacketdispersion/tree/main}
    \item \textbf{Direct link to animation}: \url{https://github.com/HugoSRagain/wavepacketdispersion/blob/main/wavepacket_dispersion.gif}
\end{itemize}

This animation illustrates the time evolution of an economic wave packet in a harmonic potential for different values of the economic action constant~$\hbar_E$. The dispersion patterns highlight the transition from quasi-classical behavior (\(\hbar_E \ll 1\)) to quantum-like dynamics (\(\hbar_E \gg 1\)).

The file can be downloaded or viewed directly in a browser.

\newpage
\section*{Appendix C: Rolling Window Estimates of $\hbar_E$ for GDP}

This appendix presents rolling window estimates of the economic action constant~$\hbar_E$ (logarithmic variant of GDP) for the four economies analyzed in Section~\ref{sec:hbarE_gdp}.  
For each country, four estimation windows are reported: $3$, $5$, $7$, and $10$ years.  
The solid curve corresponds to the rolling estimate, while the dashed line represents the full-sample (global) estimate of~$\hbar_E$.

\begin{itemize}
    \item \textbf{Data source:} World Bank, nominal GDP (current USD), annual frequency.
    \item \textbf{Countries:} France, Germany, United Kingdom, United States.
    \item \textbf{Method:} see Section~\ref{sec:method} for the definition of $\hbar_E$ and the rolling estimation procedure.
\end{itemize}

\begin{figure}[H]
    \centering
    \includegraphics[width=0.95\linewidth]{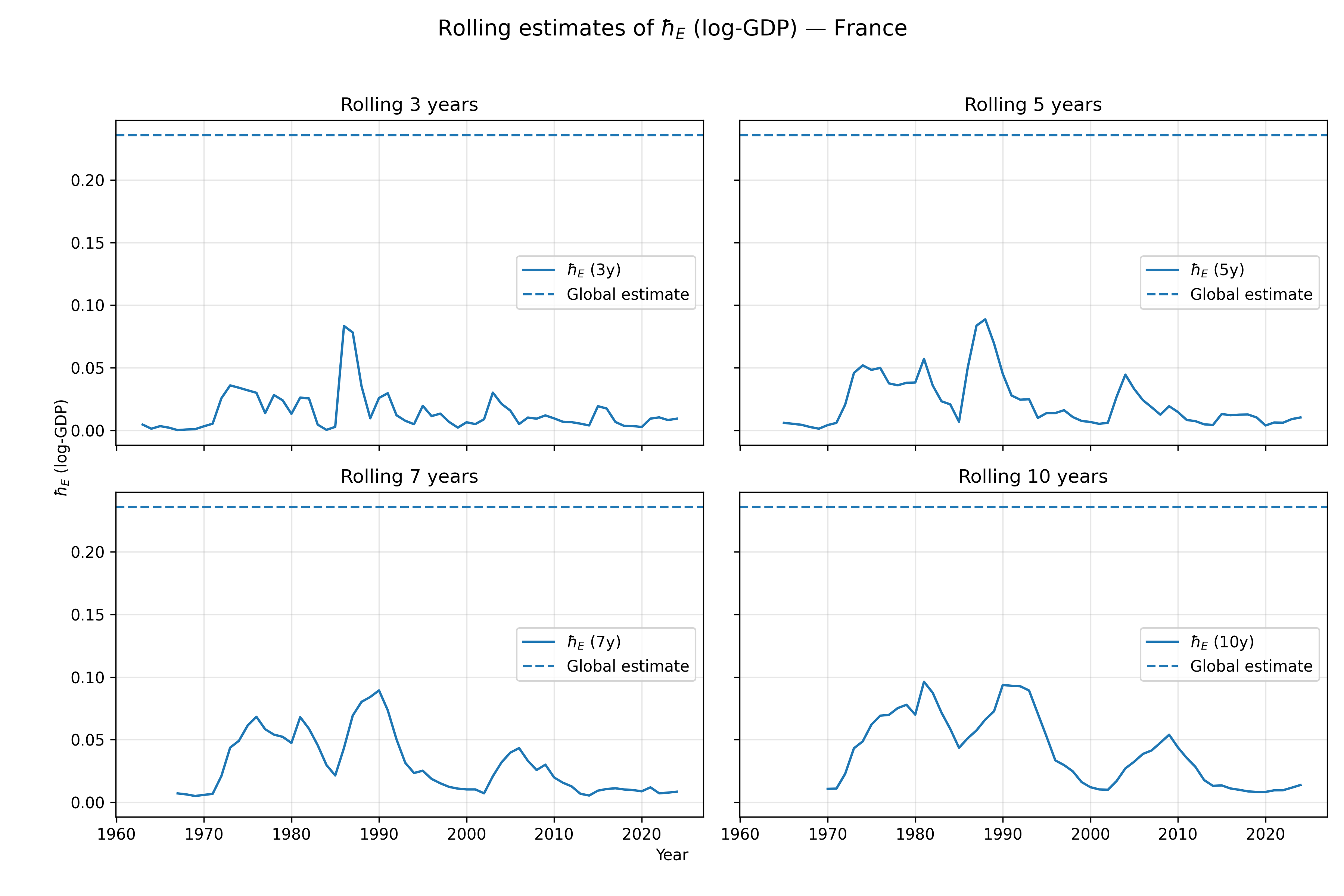}
    \caption{\textbf{France:} rolling window estimates of $\hbar_E$ (log-GDP) for windows of 3, 5, 7, and 10 years.}
    \label{fig:hbarE_France_4windows}
\end{figure}

\begin{figure}[H]
    \centering
    \includegraphics[width=0.95\linewidth]{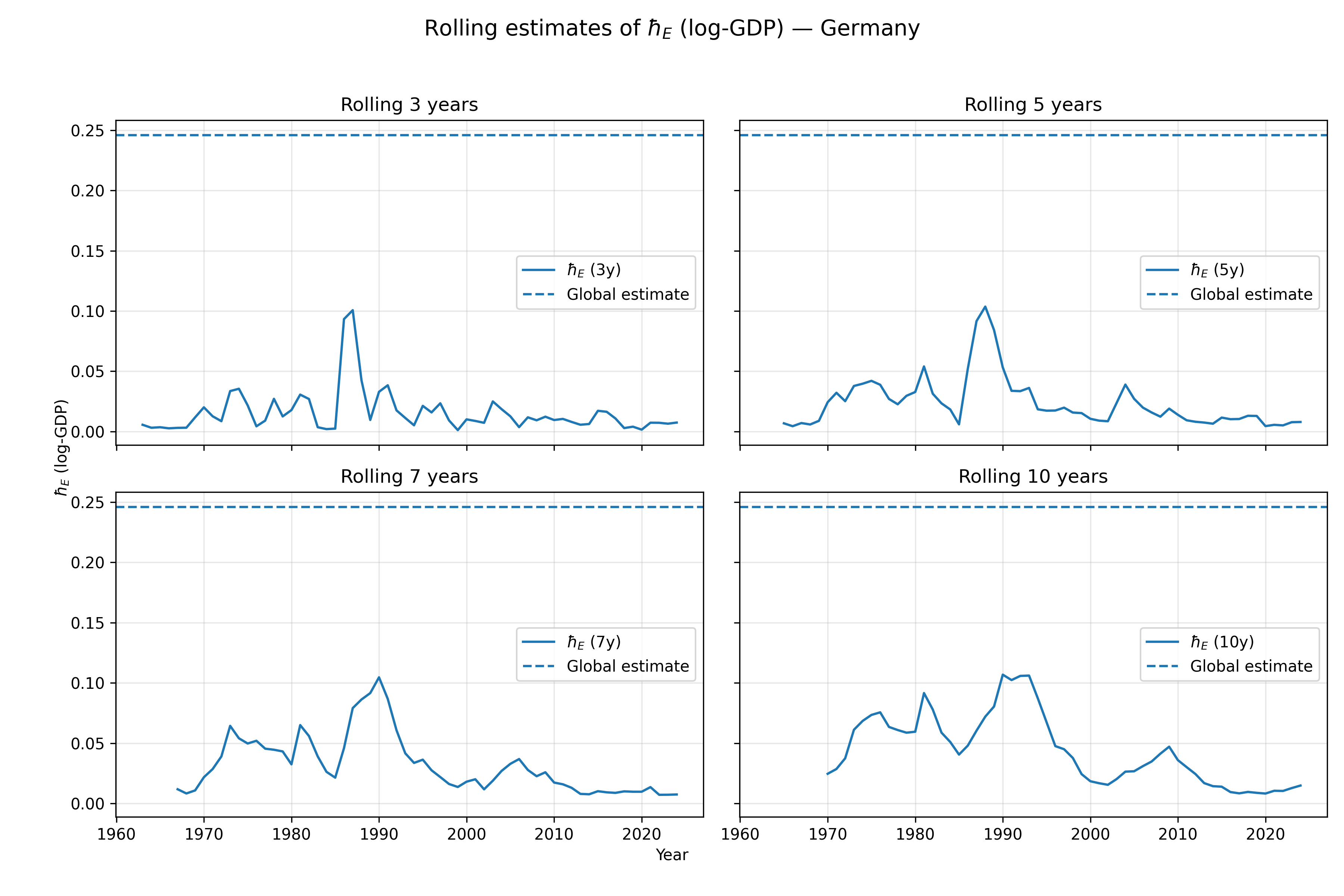}
    \caption{\textbf{Germany:} rolling window estimates of $\hbar_E$ (log-GDP) for windows of 3, 5, 7, and 10 years.}
    \label{fig:hbarE_Germany_4windows}
\end{figure}

\begin{figure}[H]
    \centering
    \includegraphics[width=0.95\linewidth]{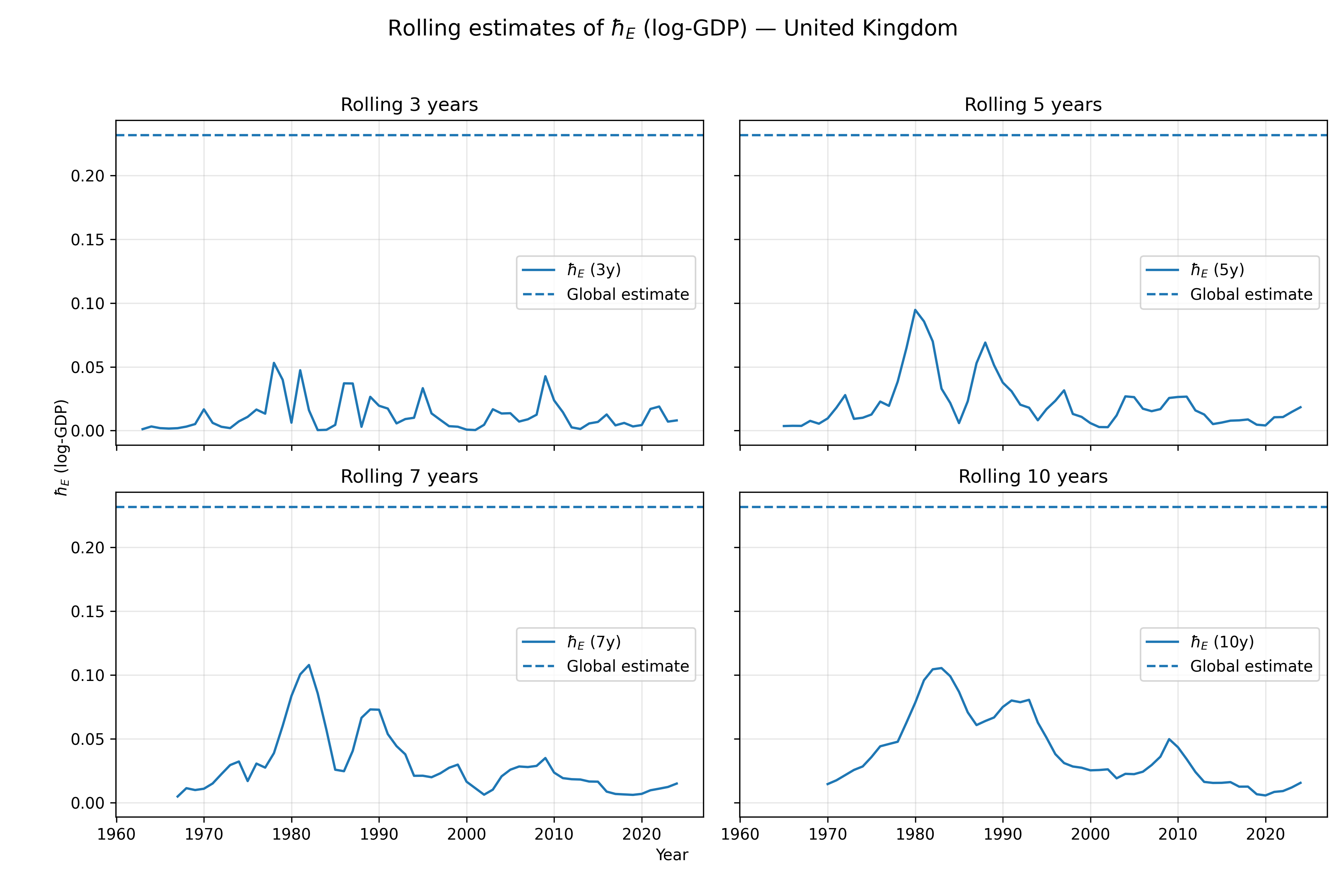}
    \caption{\textbf{United Kingdom:} rolling window estimates of $\hbar_E$ (log-GDP) for windows of 3, 5, 7, and 10 years.}
    \label{fig:hbarE_UK_4windows}
\end{figure}

\begin{figure}[H]
    \centering
    \includegraphics[width=0.95\linewidth]{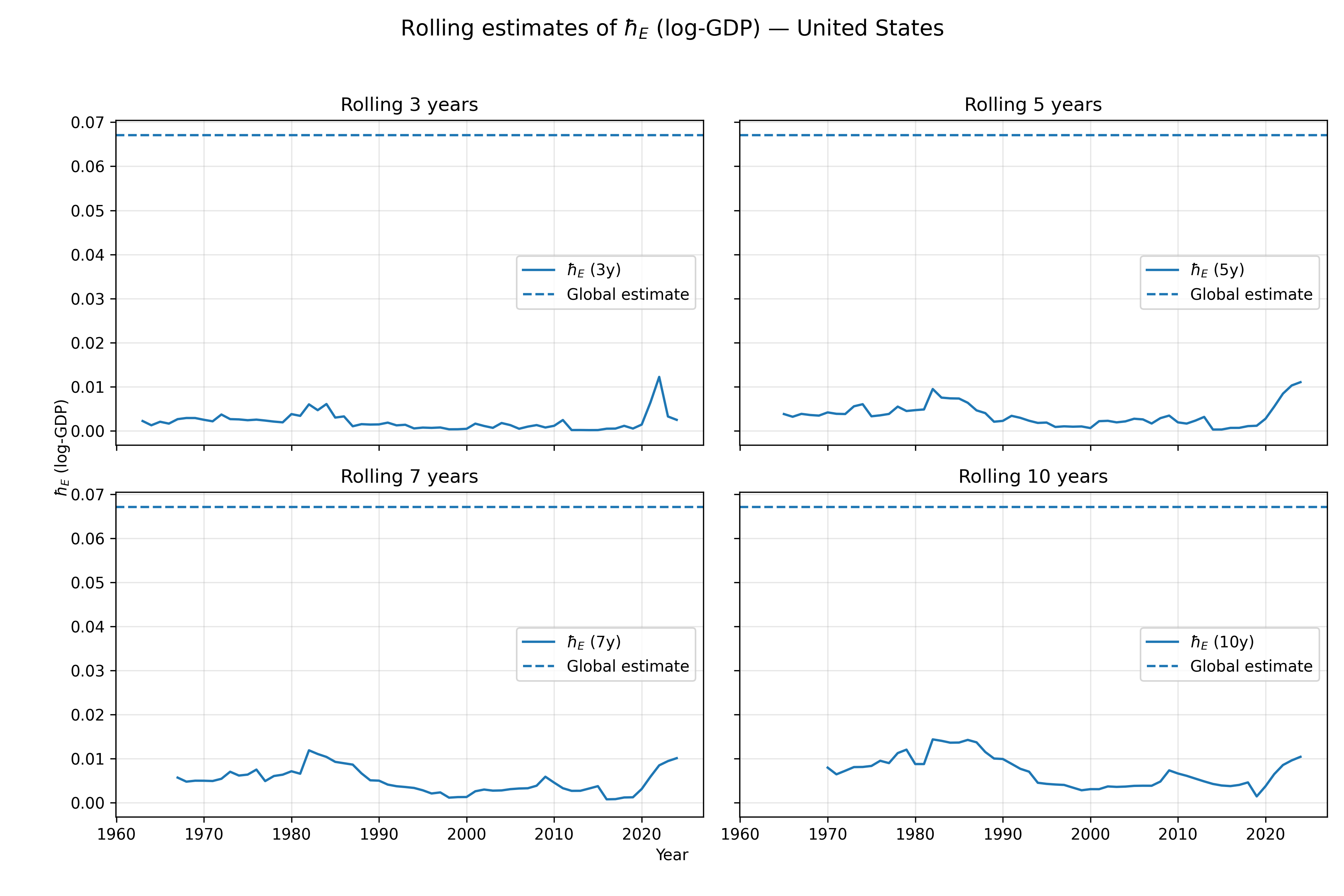}
    \caption{\textbf{United States:} rolling window estimates of $\hbar_E$ (log-GDP) for windows of 3, 5, 7, and 10 years.}
    \label{fig:hbarE_USA_4windows}
\end{figure}

\vspace{1em}
\newpage
\textbf{References :}
\begin{itemize}
  \item Planck, M. (1901). \emph{On the Law of Distribution of Energy in the Normal Spectrum}. Annalen der Physik.
  \item Einstein, A. (1905). \emph{On a heuristic point of view about the creation and conversion of light}. Annalen der Physik.
  \item Heisenberg, W. (1927). \emph{Über den anschaulichen Inhalt der quantentheoretischen Kinematik und Mechanik}. Zeitschrift für Physik.
  \item Schrödinger, E. (1926). \emph{Quantisierung als Eigenwertproblem}. Annalen der Physik.
  \item Dirac, P. A. M. (1930). \emph{The Principles of Quantum Mechanics}. Oxford University Press.
  \item von Neumann, J. (1932). \emph{Mathematische Grundlagen der Quantenmechanik}. Springer.
  \item Feynman, R. P. (1948). \emph{Space-Time Approach to Non-Relativistic Quantum Mechanics}. Reviews of Modern Physics.
  \item Cohen-Tannoudji, C., Diu, B., Laloë, F. (1977). \emph{Mécanique Quantique}. Hermann.
  \item Schulman, L. S. (1981). \emph{Techniques and Applications of Path Integration}. Wiley.
  \item Kleinert, H. (2009). \emph{Path Integrals in Quantum Mechanics, Statistics, Polymer Physics, and Financial Markets}. World Scientific.
  \item Zeh, H. D. (2007). \emph{The Physical Basis of the Direction of Time}. Springer.
  \item Preskill, J. (2015). \emph{Lecture Notes on Quantum Computation}. Caltech.
  \item Baaquie, B. E. (2007). \emph{Quantum Finance: Path Integrals and Hamiltonians for Options and Interest Rates}. Cambridge University Press.
  \item Haven, E., \& Khrennikov, A. (2013). \emph{Quantum Social Science}. Cambridge University Press.
  \item Bagarello, F. (2019). \emph{Quantum Concepts in the Social, Ecological and Biological Sciences}. Cambridge University Press.
  \item Khrennikov, A. (2010). \emph{Ubiquitous Quantum Structure: From Psychology to Finance}. Springer.
  \item Zwirn, H. (2016). \emph{La Décohérence et le problème de la mesure en mécanique quantique}. Vuibert.
  \item Aerts, D. \& Sozzo, S. (2014). Quantum Entanglement in Concept Combinations. \emph{International Journal of Theoretical Physics}, 53(10), 3587–3603.
  \item Haven, E. (2005). Pilot-wave theory and financial option pricing. \emph{International Journal of Theoretical Physics}, 44(11), 1957–1962.
  \item Wendt, A. (2015). \emph{Quantum Mind and Social Science: Unifying Physical and Social Ontology}. Cambridge University Press.
  \item Trefethen, L. N., \& Bau, D. (1997). \textit{Numerical Linear Algebra}. SIAM.
  \item Davies, E. B. (1995). \textit{Spectral Theory and Differential Operators}. Cambridge University Press.
  \item Spring-Ragain, H. (2025). \textit{Adaptation of Quantum Models to Economic Growth Theories}. SSRN Electronic Journal. Disponible en ligne : \url{https://papers.ssrn.com/sol3/papers.cfm?abstract_id=5214891}.
  \item Evans, G. W., \& Honkapohja, S. (2001). \emph{Learning and Expectations in Macroeconomics}. Princeton University Press.
  \item An, S., \& Schorfheide, F. (2007). Bayesian analysis of DSGE models. \emph{Econometric Reviews}, 26(2-4), 113–172.
  \item Sims, C. A. (2003). Implications of rational inattention. \emph{Journal of Monetary Economics}, 50(3), 665–690.
  \item Arthur, W. B. (1999). Complexity and the Economy. \emph{Science}, 284(5411), 107–109.
  \item Kirman, A. (2011). \emph{Complex Economics: Individual and Collective Rationality}. Routledge.
  \item Farmer, J. D., \& Foley, D. (2009). The economy needs agent-based modelling. \emph{Nature}, 460(7256), 685–686.
  \item Lux, T., \& Marchesi, M. (1999). Scaling and criticality in a stochastic multi-agent model of a financial market. \emph{Nature}, 397(6719), 498–500.
  \item Breuer, H.-P., \& Petruccione, F. (2002). \emph{The Theory of Open Quantum Systems}. Oxford University Press.
  \item Black, F. (1976). \emph{Noise}. The Journal of Finance, 41(3), 528-543.
  \item Campbell, J. Y., Lo, A. W., \& MacKinlay, A. C. (1997). \emph{The Econometrics of Financial Markets}. Princeton University Press.
  \item Lo, A. W., \& MacKinlay, A. C. (1988). Stock market prices do not follow random walks: Evidence from a simple specification test. The Review of Financial Studies, 1(1), 41-66.
  \item Michaely, R., Thaler, R. H., \& Womack, K. L. (1995). Price reactions around dividend announcements: Evidence from real estate investment trusts. Journal of Financial Economics, 38(2), 163-199.
  \item Roll, R. (1988). R$^2$. The Journal of Finance, 43(3), 541-566.
\end{itemize}

\end{document}